\documentclass[twocolumn,groupedaddress,showpacs,preprintnumbers,
amsmath,amssymb,aps,pre,floatfix]{revtex4-2}

\usepackage[dvipsnames]{xcolor}
\usepackage{layouts} 
\usepackage{listings}
\usepackage{dcolumn}
\usepackage{graphicx}
\usepackage{float} 

\usepackage[caption=false]{subfig} 

\usepackage{amsmath}

\begin{document}


\title{On the growing length scale in a replica-coupled glassforming liquid}


\author{Niklas Küchler}
\affiliation{Institut für Theoretische Physik II: Weiche Materie, 	Heinrich-Heine-Universität Düsseldorf,
	Universitätsstraße 1, 40225 Düsseldorf, Germany}

\author{Jürgen Horbach}
\affiliation{Institut für Theoretische Physik II: Weiche Materie, 	Heinrich-Heine-Universität Düsseldorf,
	Universitätsstraße 1, 40225 Düsseldorf, Germany}


\date{\today}

\begin{abstract}
Computer simulations are used to study a three-dimensional polydisperse model glassformer in a replica-coupling setup where an attractive field $\propto - \varepsilon Q$ of strength $\varepsilon$ can adjust the similarity of the system to a fixed reference configuration with the overlap parameter $Q$. The polydispersity in the model enables the efficient use of swap Monte Carlo in combination with molecular-dynamics simulation from which we obtain fully equilibrated liquid configurations at very low temperature, i.e., far below the critical temperature of mode-coupling theory, $T_{\rm MCT}$. When the $\varepsilon$-field is switched on, the fast dynamics with swaps allow relaxation to the stationary state at temperatures below $T_{\rm MCT}$. In the stationary state, the overlap $Q$ has a finite value that increases with increasing $\varepsilon$. For a given temperature $T$, fluctuations of the overlap around the average value become maximal at a critical field strength $\varepsilon^\star(T)$. With decreasing $T$ along this $\varepsilon^\star(T)$-line, overlap fluctuations increase and a transition from a unimodal overlap distribution to a bimodal shape occurs. We give evidence that these bimodal distributions are not due to first-order phase transitions. However, they reflect finite-size effects due to a rapidly growing length scale with decreasing temperature. We discuss the significance of this length scale for the understanding of the glass transition.

\end{abstract}


\maketitle

\section{Introduction}
At very low temperatures, the dynamics of dense liquids is intimately related to the cage effect \cite{Binder2011}. In an amorphous solid, particles are localized in cages that are formed by the surrounding particles. They are kinetically trapped in a certain (microscopically disordered) configuration, where they perform thermal motion around quasi-equilibrium positions. Caging manifests in plateau-like regions in time-dependent correlations functions, such as the overlap $Q(t)$ defined blow, which measure the similarity between a configuration at time $t$ and that at time zero. However, the cage effect is observed only on a finite timescale $\tau_{\rm lt}$ which characterizes the lifetime of an amorphous solid~\cite{lamp2022}. With increasing temperature, the lifetime $\tau_{\rm lt}$ becomes drastically smaller until the plateau-like region apparently disappears. The temperature around which this crossover from liquid- to solid-like dynamics occurs can be identified with the critical temperature of mode-coupling theory \cite{gotze2009complex}, $T_{\rm MCT}$. The overlap $Q$ can be seen as a simple order parameter that distinguishes an amorphous solid ($Q(t) \sim 1$ for $t < \tau_{\rm lt}$) from a liquid state ($Q(t) \sim 0$ for $t$ larger than a microscopic timescale) \cite{cavagna2009}.

A natural question that has been debated for a long time is whether the increasing timescale $\tau_{\rm lt}$ with decreasing temperature is accompanied by an increasing length scale \cite{adam1965temperature,Biroli2004}. Different types of length scales have been proposed and related to the drastic slowing down of liquid dynamics toward the glass transition: A dynamic length scale can be extracted from dynamic four-point susceptibilities that describe fluctuations around the average dynamics, as described in terms of intermediate scattering functions \cite{Biroli2004}. This length scale typically only shows a mild increase of a factor of $\sim 7$ when approaching the glass transition \cite{flenner2014universal}. Other length scales have been defined for systems where particles are pinned to fixed positions. For example, particles can be frozen to form a planar wall or a spherical cavity. With correlation functions that detect how unpinned particles are affected by the pinned ones (as a function of distance), one can define dynamic \cite{scheidler2004relaxation} as well as static (e.g., point-to-set) lengths \cite{montanari2006rigorous}. In computer simulations, static lengths were shown to grow by a factor of $2.5$--7 \cite{cavagna2007mosaic,biroli2008thermodynamic,biroli2013comparison,karmakar2015length}.
In experiments of colloids, pinning can be realized with holographic optical tweezers~\cite{gokhale2014growing}. In this manner, a point-to-set length was shown to reach a value of 5 diameters \cite{hima2015direct}. 
In experiments of molecular glassforming liquids,
only a small growth of static lengths between $1.4$--$1.7$ units were measured indirectly with a fifth-order dielectric susceptibility \cite{albert2016fifth} or $1.2$--$1.6$ with a soft-pinning technique \cite{das2023soft}.

An alternative way to extract static length scales for glassforming liquids is replica coupling, a setup proposed by Franz and Parisi \cite{franz1997phase, franz1998effective, Franz2013}: 
Consider a fixed reference configuration ${\bf r}_0$ that is a representative for the bulk liquid, sampled from the Boltzmann distribution at a given temperature. Another liquid is attractively coupled to ${\bf r}_0$ by introducing potential wells around each particle of the reference configuration ${\bf r}_0$. Each potential well has a microscopic spatial extent and a depth of $\varepsilon$ that controls the strength of the coupling. In the Hamiltonian of the constrained liquid, this attractive coupling is realized by introducing a term proportional to $- \varepsilon Q$ where $Q$ is the overlap between the two configurations.

Recent simulation studies have investigated replica-coupled liquids for polydisperse model glassformers \cite{RFIM_in_glassforming_liquid,guiselin2020overlap,guiselin2022statistical}, considering fully equilibrated liquids at very low temperature, i.e., at temperatures far below the mode-coupling temperature $T_{\rm MCT}$. Equilibration of liquids at such low temperature has been made possible with the use of swap Monte Carlo (SMC) in combination with molecular dynamics (MD) \cite{ninarello2017, berthier2019efficient, Kuchler2023}. In this work, hybrid MD-SMC simulations of a polydisperse model glassformer are used to investigate replica-coupled liquids for $T\ll T_{\rm MCT}$. We address the question about a static correlation length in replica coupling and its meaning for the understanding of the kinetic glass transition.   

Recent studies \cite{berthier2015evidence,RFIM_in_glassforming_liquid,guiselin2022statistical} have interpreted the behavior of replica-coupled glassforming liquids in terms of the random-field Ising model (RFIM). However, in RFIM the random field tends to destroy the long-ranged ferromagnetic order, while in our case an ``ordered'' state with a high overlap is more and more stabilized with increasing field strength $\varepsilon$. In the thermodynamic limit, a ferromagnetic phase in the RFIM is only found (for spatial dimensions $d>2$) at low temperature, provided that the random field is {\it sufficiently small}. As first predicted by Imry and Ma \cite{Imry1975}, for large random fields the system develops into a frozen domain state. Later, simulations have shown that this state is associated with a fractal structure of the domains \cite{Esser1997}. Unlike the RFIM, in the low-temperature liquid there is no phase transition in the absence of the $\varepsilon$-field: For $\varepsilon=0$ the overlap function $Q(t)$ decays from one at $t=0$ (perfect overlap) to a low value $Q_{\rm rnd} \approx 0$ for $t\to \infty$ (random overlap between two independent configurations). With increasing field strength $\varepsilon$ the average stationary-state value of the overlap steadily increases from $Q_{\rm rnd}$ toward one. On a qualitative level, a similar behavior is also found when an ideal gas is coupled to a fixed configuration.

As we shall see below from our simulation study, the interesting feature of low-temperature liquids coupled to the $\varepsilon$-field is an increase of the fluctuations around the average stationary-state value of the overlap with decreasing temperature. At a given temperature $T$, one can identify a critical value of the field strength, $\varepsilon^\star$, where these fluctuations are maximal. The analysis of susceptibilities as well as spatial correlation functions at $\varepsilon^\star$ indicates that the increase of fluctuations is accompanied by a rapidly increasing length scale, $\xi$, with decreasing temperature. This length scale measures the extent of domains of large overlap of the liquid with the reference configuration. As in similar recent simulation studies \cite{berthier2015evidence, RFIM_in_glassforming_liquid, guiselin2020overlap, guiselin2022statistical}, we observe that around $\varepsilon^\star$ the probability distribution of the overlap $Q$ becomes bimodal at low temperatures. In previous studies, this bimodal distribution has been interpreted in terms of a first-order phase transition, indicating a coexistence of a ``phase'' of high overlap with one of low overlap. Moreover, the temperature at which the crossover from a unimodal to a bimodal distribution occurs has been interpreted as a critical temperature with the critical behavior belonging to the universality class of the RFIM. Below, we propose a different interpretation of our data. We give evidence that the crossover toward a bimodal distribution is a finite-size effect: With decreasing temperature, the length scale $\xi$ becomes first of the order and then much larger than the linear dimension of the simulation box and this is the reason for the bimodality of the overlap distribution. Thus, for sufficiently large systems one would always expect a unimodal distribution. We discuss the meaning of the length scale $\xi$, especially its similarity to the point-to-set length scale.

The rest of the paper is organized as follows: In Sec.~\ref{sec:model_and_simulation_details}, we introduce the model and the main simulation details. Then, in Sec.~\ref{sec:relaxation_dynamics}, the time dependence of the overlap function in the presence of the $\varepsilon$-field is studied. A thermodynamic analysis of the overlap distributions and their fluctuations (in the stationary state) is presented in Sec.~\ref{sec:thermodynamic_analysis}. Here we compare distributions obtained with the umbrella-sampling technique with those obtained via histograms. Finite-size effects of the distributions and fluctuations are discussed. Finally, we interpret the results and draw conclusions in Sec.~\ref{sec:conclusions}.
In the appendices, we first present the calculations of the random-overlap value $Q_{\rm rnd}$ (App.~\ref{app:random_overlap}). Furthermore, we derive an exact analytical solution for the overlap distribution of the ideal gas in the presence of the external field (App.~\ref{app:ideal_gas}), providing a reference system for the liquid at high temperature. The details of the umbrella-sampling technique are given (App.~\ref{app:umbrella_sampling_technique}).
Finally, a relation between the disconnected and the connected susceptibility is derived (App.~\ref{app:relation_between_chis}).


\section{Model \& Simulation Details
\label{sec:model_and_simulation_details}}
\subsection{Model}
We consider the size-polydisperse glassformer model introduced by Ninarello~\textit{et al.}~\cite{ninarello2017} using a deterministic diameter choice as recently proposed by us~\cite{kuchler2022choice}: $N$ soft spheres with identical masses $m$ but varying ``diameters'' $\sigma_i$ labeled by particle numbers $i=1,\dots,N$ are placed in a cubic box with periodic boundary conditions. Their positions and momenta are denoted by vectors ${\bf r}_i$ and ${\bf p}_i$, respectively, and velocity ${\bf v}_i = {\bf p}_i/m$. The Hamilton function of the model is $H = K + U$ with kinetic energy $K = \sum_i^N {\bf p}_i^2/m$ and potential energy $U =  \sum_{i=1}^{N-1} \sum_{j>i}^N u \left( \frac{|{\bf r}_i - {\bf r}_j|}{\sigma_{ij}} \right)$.
%
The distance $|{\bf r}_i - {\bf r}_j|$ between two particles $i$ and $j$ is scaled by a non-additive ``interaction diameter'' $\sigma_{ij} = \frac{\sigma_i + \sigma_j}{2}( 1 - 0.2|\sigma_i - \sigma_j|)$ in the argument of the pair potential $u$ for which $u(x) = u_0 (x^{-12} + c_0 + c_2 x^2 + c_4 x^4)$ when $x < x_c$ and $u(x) = 0$ otherwise.
%
%
%
The unit of energy is $u_0$. The potential is mainly repulsive, $u(x) \sim x^{-12}$, but a polynomial with coefficients $c_0 = -28/x_c^{12}$, $c_2 = 48/x_c^{14}$ and $c_4 = -21/x_c^{16}$ smoothes $u$ at the dimensionless cutoff $x_c = 1.25$.
    
The box length $L$ is dictated by a constant number density $\rho \equiv N/L^3 = \bar{\sigma}^{-3}$ with unit of length $\bar{\sigma}$ given by the expectation value of the diameter distribution
\begin{equation}
f(s) = \begin{cases}
A s^{-3}, & \sigma_\mathrm{m} \leq s \leq \sigma_\mathrm{M},\\
0, & \mathrm{otherwise.}
\end{cases} \label{eq:polydispersity_distribution}
\end{equation}
Here $\sigma_\mathrm{m}$ is the minimum and $\sigma_\mathrm{M}$ the maximum diameter. The normalization condition $\int f(s)\,ds = 1$ sets $A = 2/(\sigma_\mathrm{m}^{-2} - \sigma_\mathrm{M}^{-2})$ and $\bar{\sigma} \equiv \int s f(s)\,ds$ implies $\sigma_\mathrm{M} = \sigma_\mathrm{m}/(2\sigma_\mathrm{m}-1)$. Only one free choice $\sigma_\mathrm{m} := 0.725$ is made so that $\sigma_\mathrm{M} = 29/18 = 1.6\overline{1}$ and $A = 29/22 = 1.3\overline{18}$.
    
To implement polydispersity, the most common approach is to draw each diameter $\sigma_i$ randomly from the distribution $f(s)$. Contrary to this \textit{stochastic} choice, we assign the diameters with a \textit{deterministic} method \cite{kuchler2022choice}, which is easily implemented from the implicit definition $\int_{-\infty}^{\sigma_i} f(s)\,ds = (i-0.5)/N $ for $i=1,\dots,N$. A deterministic choice leads to significantly superior statistical properties: (i) The histogram of diameters converges faster to $f$ for $N \to \infty$, (ii) the samples do not suffer from statistical outliers and instead the most representative realization is used, and (iii) no quenched disorder in the diameters is present that otherwise dominates sample-to-sample fluctuations at low temperatures (for example, the dynamic susceptibility is greatly enhanced with the stochastic method). This is crucial for our analysis of fluctuations in the present work.

%
The model is well-suited for the application of swap Monte Carlo which uses exchanges of particles (or equivalently their diameters) to speed up simulations: While in a \textit{binary} system the acceptance probability of swaps is only $\sim 10^{-4}$ \cite{grigera2001fast}, it is increased to $\sim 10^{-1}$ in this (continuously) polydisperse model (with an ``infinite'' number of particle species). Note that a more meaningful way to measure the efficiency of diameter swaps is a diameter correlation function~\cite{Kuchler2023}. By augmenting molecular dynamics with particle swaps, structural relaxation is accelerated by  more than $10$ orders of magnitude at low temperatures~\cite{ninarello2017}. In this way, the liquid can be equilibrated down to the (numerical) glass-transition temperature $T_\mathrm{g}^{\mathrm{SMC}} \approx 0.06$. This is far below the one of pure MD, $T_\mathrm{g}^{NVE} \approx 0.11$, which is approximately the mode-coupling temperature~\cite{Kuchler2023}.

Polydispersity increases the resistance against crystallization, here down to $T_\mathrm{g}^{\mathrm{SMC}}$~\cite{ninarello2017}, a ``confusion principle'' known from experiments of metallic glasses~\cite{greer1995metallic}. Furthermore, nonadditivity $\sigma_{ij}$ opposes demixing because it favours a chemical ordering where particle neighbors have different diameters: in this manner an effective packing fraction $\sim \sum_{ij} \sigma_{ij}^3$ (defined in Ref.~\cite{kuchler2022choice}) is decreased.

\subsection{Simulation Techniques\label{sec:simulation_details}}
To simulate the glassforming liquid, we use a hybrid scheme~\cite{berthier2019efficient} which periodically alternates between molecular dynamics (MD) and swap Monte Carlo (SMC): For the MD part, the time evolution is governed by Hamilton's equations of motion which are numerically solved with the velocity form of the Verlet algorithm. A timestep $\Delta t = 0.01\,t_0$ with unit of time $t_0 = \bar{\sigma} \sqrt{m/u_0}$ is used. After every $t_\mathrm{MD} = 0.25$ simulation time, MD is paused and $N \times s$ elementary diameter swaps are attempted with swap density $s=1$. For every of those trials, we randomly select two particles $i$ and $j$ (with a similar diameter $|\sigma_i - \sigma_j| < 0.1$ to optimize swap efficiency, see the size-bias variant in Ref.~\cite{Kuchler2023}). A swap of their diameters is attempted according to the Metropolis criterion. Then MD is repeated and so forth. The swap algorithm itself acts as a thermostat~\cite{Kuchler2023}, but we also couple the system to a Lowe-Andersen thermostat~\cite{Lowe_Andersen_T_different_masses} with a frequency of $4$ and a cutoff of $x_c$. This ensures a constant temperature when the external field (defined below) is applied.   

\subsection{Equilibration Protocol\label{sec:equilibration_protocol}}
The preparation of the samples is described in detail in Ref.~\cite{Kuchler2023}. Here, we provide a brief sketch of the protocol. For different system sizes $N = 500$, $1000$, $2048$, $4000$, $8000$, $16000$ we prepare $10$--$60$ samples each. Each sample is equilibrated at many different temperatures $T$ as follows: First the $N$ diameters are assigned to the particles with the deterministic method. Initially, we melt a crystal at a very high temperature $T=5$ with a small time step $\Delta t = 10^{-3}$ for a short time $t_\mathrm{max} = 2000$, followed by a rapid cooling to $T=0.3$ for the same duration. Then, for each target temperature $T$, the liquid is equilibrated with time step $\Delta t = 10^{-2}$ during a long simulation run for $t_\mathrm{max} = 2 \times 10^{5}$. The final configurations are the starting point of our analyses below.


\section{Relaxation dynamics\\in external $\varepsilon$-field}
\label{sec:relaxation_dynamics}

The overlap function $Q$ (or $\hat{Q}$) that we will introduce below is a measure for the similarity between two fluid configurations. In Sec.~\ref{sec:thermodynamic_analysis}, we will use this scalar-valued function to probe the phase space in terms of a free energy to calculate all \textit{equilibrium} quantities of interest.  In the present section, after introducing the general notation, we use $\hat{Q}$ to analyze the \textit{dynamic response} of the liquid to an external field $\varepsilon$. For this purpose, we investigate the relaxation dynamics in terms of a time-dependent structural correlation function $Q(t)$. The $\varepsilon$-field constrains the liquid and allows to drastically increase its kinetic stability.


\textbf{Definitions.} A liquid configuration with $N$ particles is characterized by the vector ${\bf r} = ({\bf r}_1,\dots,{\bf r}_N)$, where ${\bf r}_i$ is the position of particle $i$. In the presence of the external field, such a configuration is called a \textit{constrained liquid}. A reference configuration is denoted as ${\bf r}^0 = ({\bf r}^0_1,\dots,{\bf r}^0_{N})$.  To quantify the overlap between a configuration ${\bf r}$ and a reference configuration ${\bf r}_0$, we define the (global) overlap function $\hat{Q}$ and the (local) individual particle overlap $q({\bf r}_i)$ as
%
	\begin{align}
	\hat{Q}({\bf r},\,{\bf r}^0)  = \frac{1}{N} \sum_{i=1}^{N} q({\bf r}_i), &&
	q({\bf r}_i) =  \sum_{j=1}^{N}
	\omega\left( \frac{|{\bf r}_i - {\bf r}_j^0|}{a} \right).
	\label{eq:Q_definition}
	\end{align}
Here, the ``radius'' $a$ is set to $a = 0.4$. This microscopic length scale accounts for the small thermal fluctuations of each particle around a quasi-equilibrium position in the cage formed by surrounding particles.
The window function $\omega(x)$ in the definition of $q({\bf r}_i)$ approximates a shifted Heaviside step function, $\Theta(1-x)$, as
\begin{align}
\omega(x) 	= \begin{cases}
1 - 10 x^3 + 15 x^4 - 6 x^5, & x < 1,\\
0, & x\geq 1.
\end{cases}\label{eq:overlap_function_continuous}
\end{align}
Thus, if a particle $i$ at position ${\bf r}_i$ is closer to a particle $j$ in the reference configuration than a distance $a$, we obtain $q({\bf r}_i) \approx 1$, otherwise $q({\bf r}_i) \approx 0$.  In molecular dynamics (MD) simulations, a \textit{smooth} function $\omega$ is mandatory to enable the coupling to an external field, as finite forces (and thus finite derivatives of $\omega$) are necessary. The advantage of our definition of $\omega(x)$ compared to $\omega(x) = 2^{-x^4}$ used in Ref.~\cite{RFIM_in_glassforming_liquid} is a well-defined cutoff at $x=1$: The polynomial in Eq.~(\ref{eq:overlap_function_continuous}) is the one of smallest degree which is continuous at $x=0$ and $x=1$ up to the second derivative.
%
	

\textbf{External Field.} The coupling of a liquid at position ${\bf r}$ to a reference configuration ${\bf r}^0$ can be achieved via a linear external field $\varepsilon \geq 0$ that is conjugate to the overlap parameter $\hat{Q}$.  Thus, we can define the Hamiltonian for a constrained liquid as
\begin{equation}
	H_\varepsilon({\bf r}|{\bf r}^0) = H_0({\bf r}) - \varepsilon N \hat{Q}({\bf r},{\bf r}^0) \, ,
 \label{eq:H_epsilon_definition}
\end{equation}
where $H_0$ is the unperturbed Hamiltonian for $\varepsilon = 0$.  The coupling field $\varepsilon$ is an intensive parameter that can be illustrated as a field that introduces potential wells of depth $\varepsilon$ around each particle of the reference configuration. Note that the fixed reference configuration ${\bf r}^0$ imposes quenched disorder in the Hamiltonian.

The diameters $\sigma$ of our polydisperse model are constructed with a deterministic method, see Sec.~\ref{sec:model_and_simulation_details}. If one drew the diameters randomly from the target distribution, then an additional disorder would be present in (\ref{eq:H_epsilon_definition}), $H_\varepsilon = H_\varepsilon({\bf r}|{\bf r}^0, \sigma, \sigma^0)$. This would induce additional sample-to-sample fluctuations, as analyzed in Ref.~\cite{kuchler2022choice}.


%
\begin{figure}
\centering
\includegraphics{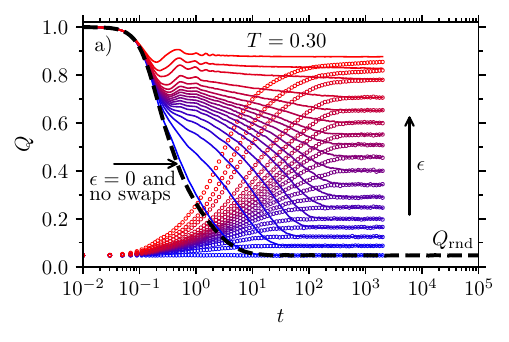}\\
\includegraphics{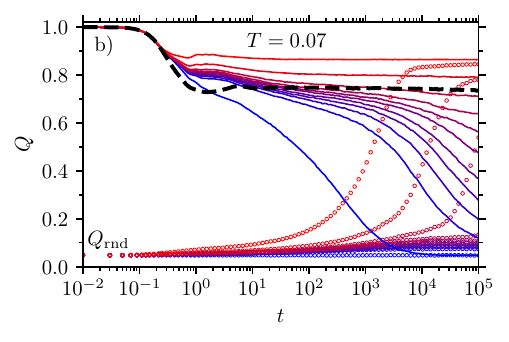}
\caption{Overlap $Q$ as a function of time $t$, averaged over $60$ samples, for a) the high temperature $T=0.3$ and b) the low temperature $T=0.07$. In both panels, the different curves correspond to different values of $\varepsilon$. With increasing field strength $\varepsilon$, the colours change from blue to red; the lowest value is $\varepsilon=0$. For both temperatures, the black dashed line corresponds to equilibrium MD dynamics \textit{without} diameter swaps. The results were obtained for $N=2048$ particles. \label{fig:Q_t_variousE}}
\end{figure}

\textit{Structural correlation function $Q(t)$.} Time-dependent overlap correlations of a liquid can be quantified in terms of the function $Q(t) = \hat{Q}({\bf r}(t), {\bf r}^0)$. In the following, ${\bf r}^0$ and ${\bf r}(0)$ correspond to equilibrated configurations at the same temperature $T$. For the dynamics of the liquid, the hybrid MD-SMC algorithm~\cite{Kuchler2023}, as described in detail above, is used.  We compare two different protocols: First, we choose ${\bf r}^0 = {\bf r}(0)$, i.e.~the reference configuration equals the initial configuration and thus $Q(t=0) = 1$ is obtained. In the second protocol, we choose an \textit{independent} reference configuration. In this case, the initial overlap at time $t=0$ is $Q(t=0) = Q_\mathrm{rnd}$, where $Q_\mathrm{rnd}$ corresponds to the value of a random overlap,
%
\begin{align}
 Q_\mathrm{rnd} \approx 0.048 > 0\, . 
 \label{eq:Q_0}
\end{align}
A precise calculation of $Q_\mathrm{rnd}$ is given in App.~\ref{app:random_overlap}.
For a binary window function, $\omega(x) = \Theta(1-x)$, the random overlap $Q_\mathrm{rnd}$ denotes the volume fraction of regions with overlap ($N$ spheres of radius $a$) and the whole simulation box.
The comparison of both protocols above allows to identify parameters $\varepsilon$ and $T$ for which a stationary state cannot be reached within the limited simulation time.

     

In Fig.~\ref{fig:Q_t_variousE}, we show $Q$ as a function of time $t$, averaged over $60$ samples, for both kinds of protocols and different $\varepsilon$. The value of $\varepsilon$ increases from blue to red. We also show pure MD dynamics (without diameter swaps) for the case $\varepsilon=0$ (black dashed line). Figures~\ref{fig:Q_t_variousE}a ($T=0.30$) and~\ref{fig:Q_t_variousE}b ($T=0.07$) show results for two \textit{qualitatively different} temperature regimes.


%
\begin{figure}
\centering
\includegraphics{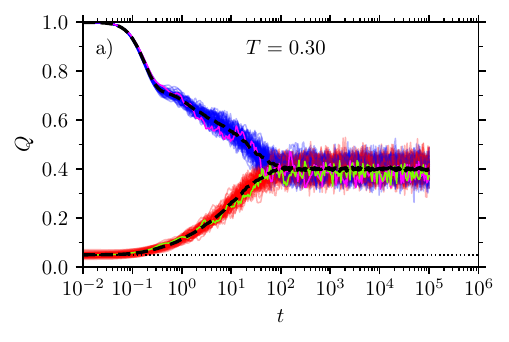}\\
\includegraphics{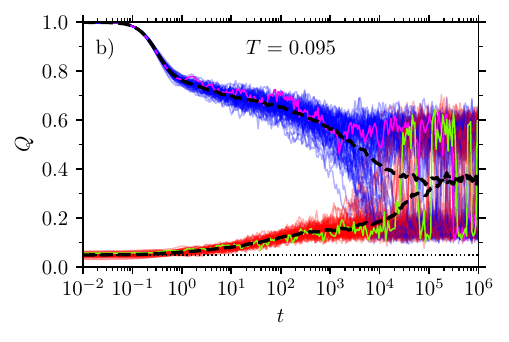}\\
\includegraphics{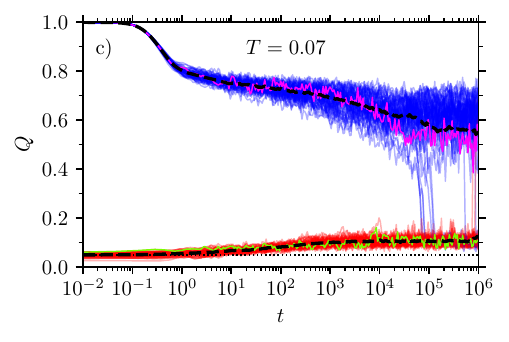}
\caption{Overlap $Q$ as a function of time $t$ for $60$ individual samples at a given temperature $T$ and at $\varepsilon=\varepsilon^\star(T)$. a) High $T=0.30$, b) intermediate $T=0.095$, c) low $T=0.07$. A single trajectory is highlighted for the protocol with $Q(t=0) = 1$ in pink and for $Q(t=0)=Q_\mathrm{rnd}$ in green. Results for systems with $N=2048$ particles are shown. \label{fig:Q_t_individual_samples}}
\end{figure}

\textit{$\varepsilon = 0$.}
That the temperature choices above correspond to different regimes can be best understood in the absence of the external field, $\varepsilon=0$, and for the first protocol, $Q(0)=1$.  For both temperatures, $Q(t)$ decays to the random overlap $Q_\mathrm{rnd}$ as the particles completely decorrelate from their initial positions ${\bf r}(0)$ due to their thermal motion. For the high temperature $T=0.30$, the decay of $Q(t)$ is monotonic, revealing a fluid-like state where particles diffuse after a relatively short time. In contrast, for the very low temperature $T=0.07$ simulated \textit{without} diameter swaps ($NVE$ dynamics, see the black dashed line), a plateau is stable for a \textit{long} timescale during which each particle is trapped within a cage formed by its neighbors. This phenomenology characterizes $T=0.07$ as an amorphous solid.  For the same $T=0.07$ but \textit{with hybrid MD-SMC dynamics featuring diameter swaps}, the particles decorrelate on a much shorter (computationally accessible) timescale $\approx 10^3$. This allows to fully equilibrate the fluid at such a low $T$ using the procedure described in Sec.~\ref{sec:equilibration_protocol}.  For the second protocol, it is $Q(t) = Q_\mathrm{rnd}$ for all $t$, as the constrained liquid is independent from the reference one right from the start.


\textit{$\varepsilon > 0$.}
The case $\varepsilon > 0$ is qualitatively different. Here, particles favor positions with higher overlap, as these are associated with potential wells. Two obvious observations are made upon increasing $\varepsilon$: Firstly, the stationary state has an increasing overlap value, $Q(t\to\infty) > Q_\mathrm{rnd}$. This behavior is also found for the ideal gas (i.e., a liquid with pair potential $U \equiv 0$) for which we explicitly calculate the stationary-state value as a function of $\varepsilon$ in App.~\ref{app:ideal_gas}.
Secondly, relaxation times onto this longtime plateau increase -- in the case of the lower temperature $T=0.07$, this increase is drastic. By increasing $\varepsilon$, relaxation times eventually reach a maximum and then decrease again. Here, the constrained liquid is trapped close to the reference configuration, associated with a high value of the overlap, where it remains forever. For the first protocol with $Q(0)=1$, this behaviour is reminiscent of the cage effect, but occurs even at a high temperature $T=0.30$ where the bulk fluid ($\varepsilon=0$) does not show a two-step decay in structural correlation functions. Overall, the phenomenology upon increasing $\varepsilon$ can be described as an increasing kinetic stability.


At a high $T=0.30$ and any given $\varepsilon$, both protocols that we analyze converge to the same stationary value of $Q$ within our simulation time-window. In contrast, for $T=0.07$, a range of $\varepsilon$ values exist for which the relaxation times become too large; the stationary state cannot be reached for $t \leq 10^5$. In other words, we cannot measure equilibrium properties of the system here.  To circumvent this problem, we will use an importance-sampling technique called umbrella sampling for a thermodynamic analysis in the next section.  We will show that both methods yield the same results at sufficiently high temperatures, where the direct approach of the present section still works (i.e., reaches equilibrium). Thereby, we demonstrate that the overlap distributions obtained via umbrella sampling can be directly interpreted in terms of the direct kinetic approach, discussed in this section.



For our analysis in the next Sec.~\ref{sec:thermodynamic_analysis}, we are especially interested in an intermediate value $\varepsilon = \varepsilon^\star(T)$, for which fluctuations are maximal, see the definition (\ref{eq:critical_epsilon}) below. Here $Q(t\to\infty) \approx 0.36$.  In Fig.~\ref{fig:Q_t_individual_samples}, we show many individual trajectories $Q(t)$ for this $\varepsilon^\star(T)$ for three different temperatures $T$.  For the protocol with $Q(t=0) = 1$, we show $59$ trajectories in blue and one in pink, for the protocol with $Q(t=0)=Q_\mathrm{rnd}$, we illustrate $59$ trajectories in red and one in green. For both protocols, an average curve over the corresponding $60$ samples is shown as a black dashed line. 

We observe that with decreasing temperature $T$ the fluctuations of the trajectories $Q(t)$ around the average curve increase.  For the very low $T = 0.07$, we cannot reach the stationary state as indicated by the average curves of both protocols. Fluctuations seem to decrease, but this is only a kinetic effect, as the accessible simulation time is too short to resolve the fluctuations. In the next section, umbrella sampling allows to access equilibrium states even at very low temperatures. We will show that fluctuations increase \textit{at least} up to the glass-transition temperature $T_g^\mathrm{SMC} \approx 0.06$, thereby providing evidence against a critical point above $T_g^\mathrm{SMC}$.

%
	
%
\section{Thermodynamic Analysis
\label{sec:thermodynamic_analysis}}
\subsection{Is there a phase transition?
\label{sec_ta_a}}
Now, we analyze the thermodynamic properties of the system in the presence of the external field $\varepsilon$, based on the Hamiltonian given by Eq.~(\ref{eq:H_epsilon_definition}). The central quantity is the probability distribution of overlaps with respect to a configuration ${\bf r}^0$ in the presence of the field $\varepsilon$, $P_\varepsilon(Q|{\bf r}^0)$. At low temperatures and appropriate values of $\varepsilon$, a bimodal distribution is observed that is reminiscent of the occurrence of a (putative) first-order phase transition, associated with the coexistence of a ``phase'' with low overlap (the ``liquid phase'') with one with a large overlap (the ``glass phase''). In order to reveal whether such an interpretation is valid, we determine overlap distributions (Sec.~\ref{sec_ta_b}), susceptibilities (Sec.~\ref{sec_ta_c}), and spatial correlation-functions of local overlaps (Sec.~\ref{sec_ta_d}).

To determine the probability distributions of overlaps we face the following central problems. First, relaxation in the $\varepsilon$-field eventually becomes so slow that we cannot reach the stationary state with the ``direct'' simulation approach, as shown in Sec.~\ref{sec:relaxation_dynamics}. In this case, we cannot measure the correct equilibrium distribution via a histogram of $Q$ values within our finite simulation time. Second, when we calculate a histogram of $Q$, even when in equilibrium, $Q$ values with a low probability will suffer from immensely poor sampling. A solution to both problems is umbrella sampling~\cite{umbrella_sampling_kastner2011,guiselin2022statistical}, an importance sampling technique that we describe in appendix~\ref{app:umbrella_sampling_technique}. The method introduces a bias (harmonic potential) that forces the system to stay around a certain value of the overlap. Then, the unbiased distribution can be obtained from the biased simulations. The method comes at the cost of a high computational effort, since many simulations (covering different biases) have to be performed.

Below in Sec.~\ref{sec_ta_b}, we first discuss the probability distributions, as obtained from umbrella sampling, and compare them to distributions that we calculated directly from the long-time behavior of the overlap, presented in Sec.~\ref{sec:relaxation_dynamics}. On first sight, the transition from \textit{unimodal} to \textit{bimodal} probability distributions seem to suggest a line of first-order transitions that end in a critical point. However, a system-size analysis hints toward the possibility that bimodality is a finite-size effect.
Then, in Sec.~\ref{sec_ta_c}, we introduce thermal and disorder averages as well as connected and disconnected susceptibilities to describe fluctuations. We will see that the behavior of the susceptibilities supports the interpretation that the appearance of bimodal distributions is only a finite-size effect. In particular, this becomes evident when we consider spatial correlation functions of local overlaps in Sec.~\ref{sec_ta_d}. They indicate a growing static length scale of clusters, classified as regions of high and low overlap.

\subsection{Equilibrium distributions \label{sec_ta_b}}
We have to account for the quenched disorder introduced by the choice of the reference configuration ${\bf r}^0$ in the Hamiltonian $H_\varepsilon({\bf r}|{\bf r}^0)$. Each ${\bf r}^0$ is supposed to represent a typical fluid snapshot taken at a temperature $T_0$. For a stochastic treatment, ${\bf r}^0$ is considered to be a random variable distributed according to the canonical ensemble of the unbiased Hamiltonian $H_0$, i.e.,
%
\begin{equation}
P({\bf r}^0) \propto \exp( - H_0({\bf r}^0)/(k_\mathrm{B} T_0) ),
\label{eq:P_NVT_unbiased}
\end{equation}
with a normalization constant given by integrating over all $3 \times N$ degrees of freedom.  Similarly, for the constrained liquid ${\bf r}$ in the $\varepsilon$-field, we have $\tilde{P}_\varepsilon({\bf r} | {\bf r}^0) \propto \exp( - H_\varepsilon({\bf r}|{\bf r}^0)/(k_\mathrm{B} T))$, which is a conditional probability, accounting for the influence of ${\bf r}^0$ on ${\bf r}$.
    
The probability distribution of the overlap function $\hat{Q}$ on $[0,1]$ is inherited from $\tilde{P}_\varepsilon$ as
\begin{align}
P_\varepsilon(Q|{\bf r}^0) &= 
\frac{ \int_{\mathcal{V}} \delta\left(Q - \hat{Q}({\bf r},{\bf r}^0)\right) e^{-H_\varepsilon({\bf r}|{\bf r}^0)/(k_\mathrm{B} T)}\,\text{d}{\bf r} }
{ \int_{\mathcal{V}} e^{- H_\varepsilon({\bf r}|{\bf r}^0)/(k_\mathrm{B} T)}\,\text{d}{\bf r} }.
\label{eq:P_eps_Q_r0} 
\end{align}
%

%
\begin{figure}
\centering
\includegraphics{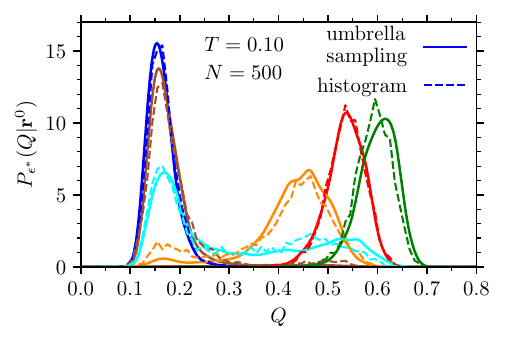}
\caption{Probability distribution $P_\varepsilon(Q|{\bf r}^0)$ for six different reference configurations ${\bf r}^0$, calculated from a histogram of $Q(t)$ data (dashed lines) or from umbrella sampling (solid lines). $\varepsilon$ is set to $\varepsilon^\star$, see definition~(\ref{eq:critical_epsilon}). \label{fig:P_Q_Ec_comparison_direct_with_umbrella}}
\end{figure}
For the ideal gas we explicitly calculate the overlap distribution $P_\varepsilon^\mathrm{id}$ in App.~\ref{app:ideal_gas}. We obtain a binomial distribution (which is similar to a Gaussian distribution for large $N$), whose single peak can be arbitrarily tuned with variation of $\varepsilon$. The variance of $P_\varepsilon^\mathrm{id}$ is maximal at a critical value $\varepsilon =\varepsilon^\star$ for which the peak is located at $Q=0.5$. While the quenched disorder ${\bf r}^0$ is not relevant for the ideal gas, we will see now that it truly is for the liquid.

In Fig.~\ref{fig:P_Q_Ec_comparison_direct_with_umbrella}, for the liquid, we show $P_\varepsilon(Q|{\bf r}^0)$ at the critical $\varepsilon =\varepsilon^\star(T)$, see the definition (\ref{eq:critical_epsilon}) below, where fluctuations are maximal.  

We compare two different approaches: The first is the ``direct'' simulation approach (dashed lines) of Sec.~\ref{sec:relaxation_dynamics}. Here, each differently colored curve $P_\varepsilon(Q|{\bf r}^0)$ was calculated as a normalized histogram of $Q(t)$ overlap data-points, each from a different simulation run at a given reference configuration ${\bf r}_0$. Only data with $t \in [2.5 \times 10^5,\,5 \times 10^5]$ are used after the external field was switched on at time $t=0$. For $T \geq 0.095$, this ensures that the data correspond to the stationary state: We have checked that (i) the two protocols with $Q(0) = 1$ and $Q(0) = Q_\mathrm{rnd}$ yield the same histograms and (ii) that the histogram does not change when the time window is shifted to later times. Note that we have used a bin width of $0.01$ for the calculation of the histograms. 

The second approach to obtain the distribution is umbrella sampling (solid lines in Fig.~\ref{fig:P_Q_Ec_comparison_direct_with_umbrella}) by calculating $P_\varepsilon(Q|{\bf r}^0) = C \exp( - [ F_0(Q|{\bf r}^0) - \varepsilon N Q ]/(k_\mathrm{B} T))$ from the free energy $F_0(Q|{\bf r}^0)$, see appendix \ref{app:umbrella_sampling_technique}. We clearly see that both approaches yield the same results, even quantitatively. Thus we can infer that the umbrella-sampling method provides the correct equilibrium distribution for $T \geq 0.095$. At lower temperatures a range of $\varepsilon$ values exist for which the direct approach does not allow to reach equilibrium -- due to the increased kinetic stability analyzed in Sec.~\ref{sec:relaxation_dynamics}. Thus, for $T < 0.095$ we can only rely on umbrella sampling. Again, to see whether the distribution is correct, we check whether the two protocols with $Q(0)=1$ or $Q(0)=Q_\mathrm{rnd}$ yield the same results.


\textbf{Disorder-averaged distribution $\overline{P}_{\varepsilon^\star}(Q)$.} For the remaining part of this section we want to analyze the disorder-averaged distribution $\overline{P}_{\varepsilon^\star}$ of the overlap $Q$ at the critical $\varepsilon^\star(T)$ for different temperatures $T$ and different particle numbers $N$. For $T \geq 0.095$, each histogram $\overline{P}_{\varepsilon^\star}$ shown below is the average over $60$ individual histograms $P_{\varepsilon^\star}(Q|{\bf r}^0)$ which were obtained with the direct approach as described in the paragraphs above. For the low temperature $T=0.07$, we used umbrella sampling to first calculate $P_{\varepsilon^\star}(Q|{\bf r}^0)$ in $10$ independent simulations and then averaged over those distributions.

\begin{figure}
\centering
\includegraphics{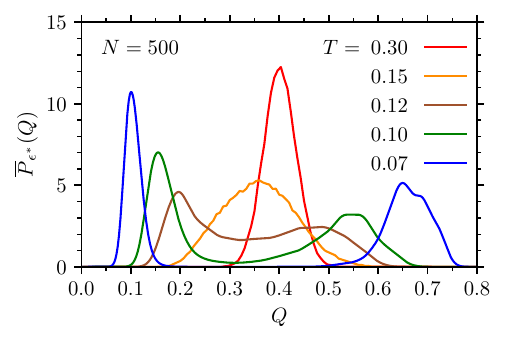}
\caption{Disorder-averaged probability distribution $\overline{P}_{\varepsilon^\star}$ of the overlap $Q$ at the critical $\varepsilon^\star$ for different temperatures $T$. \label{fig:P_Q_Ec_averaged_manyT}}
\end{figure}

In Fig.~\ref{fig:P_Q_Ec_averaged_manyT}, the average distribution $\overline{P}_{\varepsilon^\star}(Q)$ is shown for different temperatures $T$ with $N=500$ particles. A qualitative change with decreasing $T$ can be observed: 
At high temperatures $T$ the distribution is unimodal (with a single peak), as it is the case for the ideal gas, cf.~App.~\ref{app:ideal_gas}.
However, unlike the ideal gas, the distribution of the liquid becomes broader with decreasing $T$ before it transitions into a bimodal distribution (with two peaks).

Note that bimodality is not only the result of disorder-averaging, as we can infer from the individual distributions in Fig.~\ref{fig:P_Q_Ec_comparison_direct_with_umbrella}. Here two of six individual curves ${P}_{\varepsilon^\star}(Q|{\bf r}_0)$ are bimodal at $T=0.10$. However, as a consequence of disorder-averaging, $\overline{P}_{\varepsilon^\star}$ is clearly bimodal at $T=0.10$, even though most ${P}_{\varepsilon^\star}(Q|{\bf r}_0)$ are unimodal.

\begin{figure}
\centering
\caption{Disorder-averaged probability distribution $\overline{P}_{\varepsilon^\star}$ of the overlap $Q$ at the critical $\varepsilon^\star$ for different temperatures $T$ and different number of particles $N$. While a)--d) where obtained via a histogram, for e) we used umbrella sampling. \label{fig:P_Q_Ec_averaged_differentN}}
\includegraphics{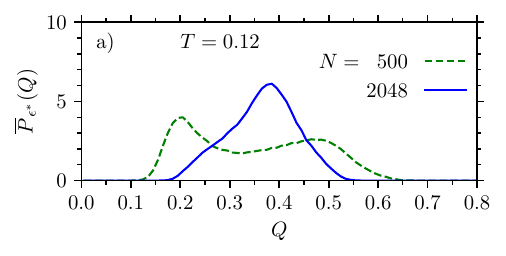}\\
\includegraphics{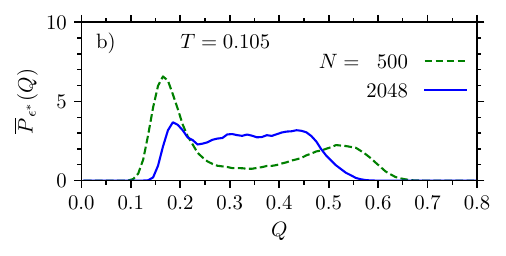}\\
\includegraphics{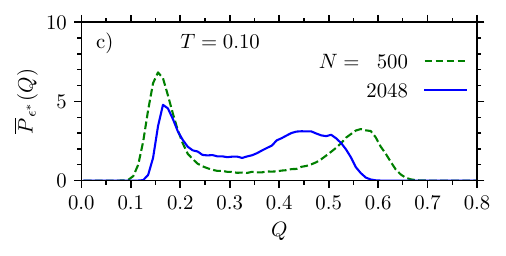}\\
\includegraphics{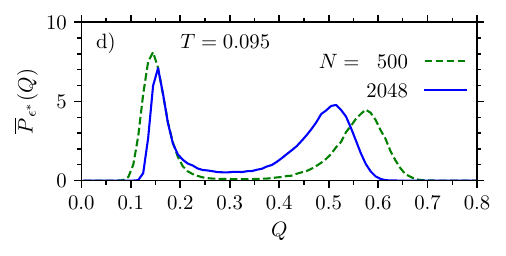}\\
\includegraphics{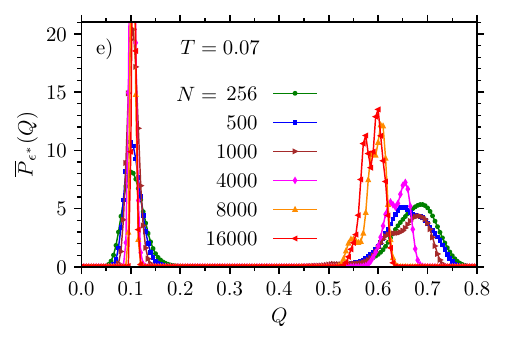}
\end{figure}

Figure \ref{fig:P_Q_Ec_averaged_differentN} compares $\overline{P}_{\varepsilon^\star}(Q)$ for different system sizes at many different temperatures $T$. In the considered temperature range $T=0.07$--$0.12$, the distribution is always bimodal for $N=500$ particles, while for $N=2048$ the transition from unimodality to bimodality with decreasing $T$ is apparent.

Bimodality has been interpreted in Refs.~\cite{berthier2015evidence,RFIM_in_glassforming_liquid,guiselin2022statistical} as indication for a phase transition between a high- and a low-overlap phase. However, our data suggest the possibility that bimodality could only be a finite-size effect: In Figs.~\ref{fig:P_Q_Ec_averaged_differentN}a--d, for fixed temperature $T$, the two peaks of the distribution tend to approach each other when $N$ is increased from $500$ to $2048$. For the very low temperature $T=0.07$ in Fig.~\ref{fig:P_Q_Ec_averaged_differentN}e, obtained via umbrella sampling, our statistics are worse. However, the high-overlap peak seems to shift to lower $Q$ values with increasing $N$ as well. A similar trend can be observed by careful inspection of the corresponding figures in Refs.~\cite{RFIM_in_glassforming_liquid,guiselin2022statistical}. To further elucidate the issue whether the bimodality of the distributions $\overline{P}_{\varepsilon^\star}(Q)$ is due to a finite-size effect or a phase transition, in the next section we study susceptibilities along the ``critical'' $\varepsilon^\star(T)$-line.
\subsection{Fluctuations \label{sec_ta_c}}
To address the question whether the fluid undergoes a phase transition, the study of fluctuations in terms of connected and disconnected susceptibilities is essential. To define these susceptibilities, we have to distinguish between a thermal average and a disorder average.
All numerical results presented in this section were obtained via umbrella sampling.

\textbf{Disorder average.} For an observable $B({\bf r}^0)$, we define the disorder average [according to Eq.~(\ref{eq:P_NVT_unbiased})] as
\begin{align}
\overline{B} = \int B( {\bf r}^0) P({\bf r}^0)~d{\bf r}^0.
\label{def:disorder_average}
\end{align}
In practice, we calculate the arithmetic mean of $B({\bf r}^0)$ among $10$ different reference configurations ${\bf r}^0$.

\textbf{Thermal average.} For an observable $A(\hat{Q})$, depending only on the overlap $\hat{Q}({\bf r},{\bf r}^0)$, the thermal average is
\begin{align}
\langle A \rangle ({\bf r}^0) = \int_0^1 A(Q) P_\varepsilon(Q|{\bf r}^0)\,dQ.
\end{align}
%
The thermal average of the overlap, $\langle \hat{Q} \rangle({\bf r}^0)$, can be identified via a time average of $Q(t)$ data points of a single simulation run with reference configuration ${\bf r}^0$, cf.~Fig.~\ref{fig:Q_t_individual_samples}, obtained from the kinetic protocol of Sec.~\ref{sec:relaxation_dynamics} after the stationary state is reached.

\begin{figure}
\centering
\includegraphics{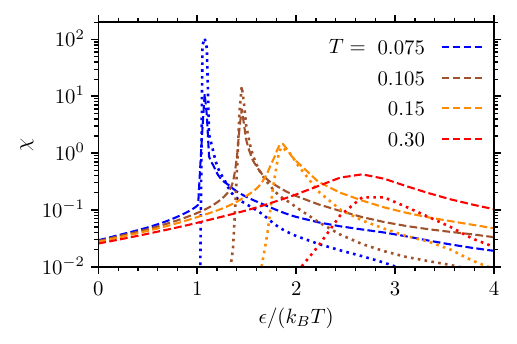}
\vspace{-0.5cm}
\caption{Connected susceptibility $\chi_\mathrm{con}$ (dashed lines) and disconnected susceptibility $\chi_\mathrm{dis}$ (dotted lines) as a function of $\varepsilon$ for various temperatures $T$. Results from umbrella sampling with $10$ independent simulations with $N=2048$ particles are shown. \label{fig:sus_eps_variousT}}
\end{figure}

\textbf{Susceptibilities}. For the overlap function $\hat{Q}$ at a given reference configuration ${\bf r}^0$, we first introduce the thermal susceptibility $\chi_\mathrm{thm}({\bf r}^0) = N [\langle \hat{Q}^2 \rangle - \langle \hat{Q} \rangle ^2]$. This quantity can be identified with the fluctuations of a selected trajectory in Fig.~\ref{fig:Q_t_individual_samples}.  Averaging over the disorder ${\bf r}^0$ yields the sample-independent
connected susceptibility
%
\begin{align}
\chi_\mathrm{con} 
= \overline{ \chi_\mathrm{thm}}
= N [\overline{\langle \hat{Q}^2 \rangle} - \overline{ \langle \hat{Q} \rangle ^2 }].	
\label{eq:chi_con}
\end{align}
The thermal mean $\langle \hat{Q} \rangle$ is a (quenched) random variable as it depends on ${\bf r}^0$. Its fluctuations are measured between samples (and are thus qualitatively different from $\chi_\mathrm{con}$) via the disconnected susceptibility
\begin{align}
\chi_\mathrm{dis} &= 
N [\overline{\langle \hat{Q} \rangle^2 } - \overline{\langle \hat{Q} \rangle}^2].
\label{eq:chi_dis}
\end{align}
It is easy to show that the variance $\mathrm{Var}(Q)$ of the disorder-averaged distribution $\overline{P}_\varepsilon$, analyzed in the previous section, is related to the susceptibilities as $N \mathrm{Var}(Q) = \chi_\mathrm{con} + \chi_\mathrm{dis}$.


In Fig.~\ref{fig:sus_eps_variousT}, we plot the connected susceptibility $\chi_\mathrm{con}$ (dashed line) and the disconnected susceptibility $\chi_\mathrm{dis}$ (dotted line) as a function of field strength $\varepsilon$ for different temperatures $T$. At a given $T$, both susceptibilities exhibit a maximum at the same ``critical'' $\varepsilon$ value,
\begin{equation}
\varepsilon^\star(T) =  \arg \max_{\varepsilon}  \{\,\chi(T) \,\}.
\label{eq:critical_epsilon}
\end{equation}
Similarly, for the ideal gas the connected susceptibility has a well-defined maximum, see App.~\ref{app:ideal_gas}, but disorder fluctuations are zero.
Unlike the ideal gas, for the liquid the height of the peak grows with decreasing $T$ and the curves $\chi(\varepsilon)$ become increasingly sharp around $\varepsilon^\star$. In this sense, the fluid becomes increasingly sensitive to the external field.

\begin{figure}
\centering
\includegraphics{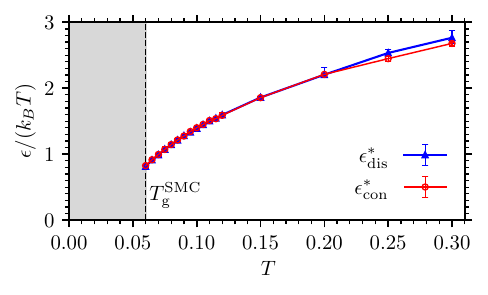}
\vspace{-0.5cm}
\caption{Critical field strength $\varepsilon^\star$ as a function of temperature $T$ for both susceptibilities, $\chi_\mathrm{con}$ and $\chi_\mathrm{dis}$. The numerical glass-transition temperature is $T_\mathrm{g}^\mathrm{SMC} \approx 0.06$ (vertical line)~\cite{kuchler2022choice}. \label{fig:Ec_T}}
\end{figure}
In Fig.~\ref{fig:Ec_T}, we plot $\varepsilon^\star$ as a function of temperature $T$ for both susceptibilities, as defined via Eq.~(\ref{eq:critical_epsilon}). Within the accuracy of our data, we find $\varepsilon^\star_\mathrm{con} \equiv \varepsilon^\star_\mathrm{dis}$. With increasing $T$ the value $\varepsilon^\star/(k_\mathrm{B}T)$ seems to saturate to a constant value, corresponding to that of the ideal gas, see App.~\ref{app:ideal_gas}.

\textbf{A critical point?}
In mean-field models of spin glasses, the critical $\varepsilon^\star(T)$-line corresponds to a line of first-order phase transitions that ends in a critical point $T_c > 0$ \cite{franz1997phase,franz1998effective}. It has been argued \cite{franz1998effective,berthier2015evidence,RFIM_in_glassforming_liquid,guiselin2022statistical} that this mean-field scenario also takes place in structural glassforming liquids. To understand if this is the case, we evaluate the two susceptibilities defined above along the critical line, i.e., we determine $\chi^\star = \chi(\varepsilon^\star)$. These values of the susceptibilities are discussed in the following.

\begin{figure}
\centering
\includegraphics{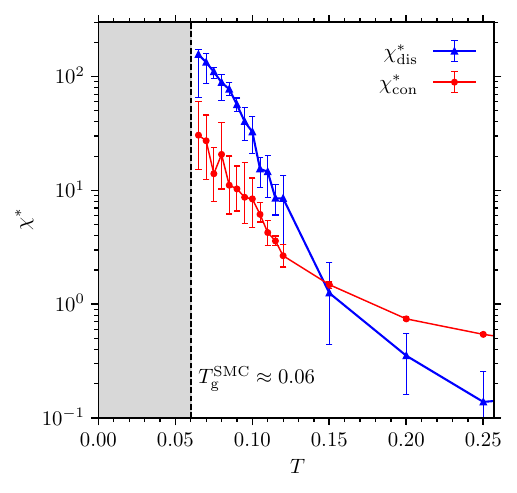}
\caption{Maximum $\chi^\star =  \max_{\varepsilon}  \{\,\chi \,\}$ of the connected susceptibility $\chi_\mathrm{con}$ (red) and the disconnected susceptibility $\chi_\mathrm{dis}$ (blue) as a function of temperature $T$. The glass-transition temperature $T_\mathrm{g}^\mathrm{SMC} \approx 0.06$ is indicated by the vertical line and the gray area. Error bars show the $90\%$ confidence interval determined via bootstrapping~\cite{efron1992bootstrap} with $100$ repetitions. Results for systems with $N=2048$ are shown. \label{fig:X_T}}
\end{figure}

Figure~\ref{fig:X_T} shows $\chi^\star_\mathrm{con}$ and $\chi^\star_\mathrm{dis}$ as a function of temperature $T$. Both susceptibilities grow upon decreasing $T$. While at high temperatures $\chi^\star_\mathrm{con} > \chi^\star_\mathrm{dis}$ holds, for low temperatures the disconnected susceptibility is larger than the connected one.


%

In general, growing thermodynamic fluctuations -- as quantified by $\chi$ -- are a typical (but not sufficient) sign for approaching a critical point. That the disconnected susceptibility grows faster than the connected one in this model has been interpreted in Ref.~\cite{RFIM_in_glassforming_liquid} as evidence for random-field Ising-model (RFIM) criticality, claiming that a critical temperature $T_c$ exists along the $\varepsilon^\star(T)$-line at $T = T_c \in [0.085, 0.10]$. However, our results are not consistent with this interpretation. As we see in Fig.~\ref{fig:X_T}, the connected susceptibility continues to grow even below $T = 0.085$, until the numerical glass-transition temperature $T_g^\mathrm{SMC} \approx 0.06$ is reached. This is evidence against the hypothesis of a critical point (at least for equilibrium temperatures, $T > T_\mathrm{g}^\mathrm{SMC}$), in which case we would expect a decrease of $\chi$ below $T_c$.

\begin{figure}
\centering
\includegraphics{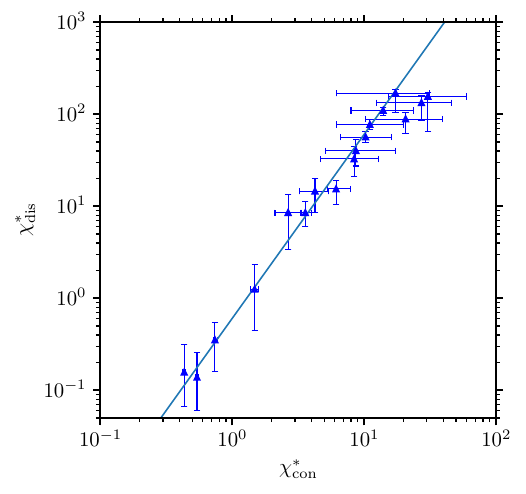}
\vspace{-0.3cm}
\caption{Disconnected susceptibility maximum, $\chi^\star_\mathrm{dis}$, versus connected susceptibility maximum, $\chi^\star_\mathrm{con}$. The data points are for a fixed number of particles $N=2048$ but different temperatures $T$ with $T_\mathrm{g}^\mathrm{SMC} \approx 0.06 \leq T \leq 0.30$. Error bars show the $90\%$ confidence interval as determined via bootstrapping~\cite{efron1992bootstrap}. \label{fig:Xdis_X_con}}
\end{figure}

Figure~\ref{fig:Xdis_X_con} shows the maximum disconnected susceptibility $\chi^\star_\mathrm{dis}$ as a function of the maximum connected susceptibility $\chi^\star_\mathrm{dis}$. The blue line indicates a proportionality $\chi^\star_\mathrm{dis} \propto (\chi^\star_\mathrm{con})^2$ which is convincingly followed by our data. This proportionality has been interpreted as evidence for RFIM criticality \cite{berthier2015evidence,RFIM_in_glassforming_liquid}. However, the equality 
\begin{equation}
    \chi^\star_\mathrm{dis} = N \mathrm{Var}(\varepsilon) (k_\mathrm{B} T)^{-2} (\chi^\star_\mathrm{con})^2
    \label{eq:chis_relation_maintext}
\end{equation}
is a \textit{generic} property of systems with quenched disorder (induced by ${\bf r}_0$ in this case). Here $\mathrm{Var}(\varepsilon)$ is the variance of disorder-specific ``critical'' $\varepsilon_0$, as defined in App.~\ref{app:relation_between_chis}. In App.~\ref{app:relation_between_chis} we derive Eq.~(\ref{eq:chis_relation_maintext}) 
under the simple physical assumption that the overlap is self-averaging.
Thus, the relation $\chi^\star_\mathrm{dis} \sim {\chi^\star_\mathrm{con}}^2$ is not a unique feature of RFIM criticality.
    
\begin{figure}
\centering
\includegraphics{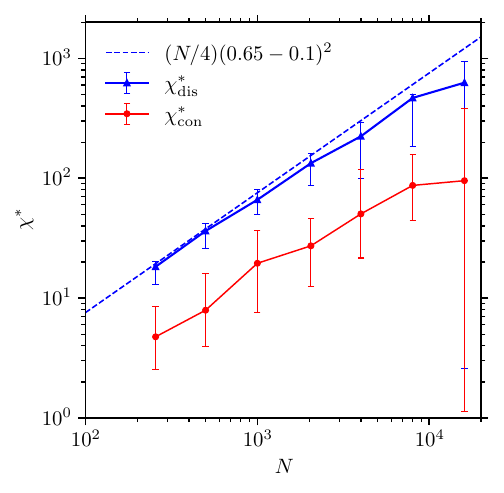}
\vspace{-0.3cm}
\caption{Maximum $\chi^\star$ of the connected susceptibility $\chi_\mathrm{con}$ (red) and disconnected susceptibility $\chi_\mathrm{dis}$ (blue) as a function of particle number $N$ for fixed temperature $T=0.07$. \label{fig:X_N}}
\end{figure}

In Fig.~\ref{fig:X_N} the connected (red) and disconnected susceptibility (blue) are shown as a function of system size $N$ for $T=0.07$. To understand the increase of $\chi^\star_\mathrm{dis}$ with increasing $N$, we first note that at such a low temperature the corresponding overlap distribution, cf.~Fig.~\ref{fig:P_Q_Ec_averaged_differentN}e, is bimodal for all considered particle numbers $N = 256$, 500, 1000, 2048, 4000, 8000, 16000. The high- and low-overlap peaks in Fig.~\ref{fig:P_Q_Ec_averaged_differentN}e are at $Q_2 \approx 0.65$ and $Q_1 \approx 0.1$, respectively. If we approximate the distribution with a Bernoulli distribution, i.e., $P(Q=Q_1) = 0.5$ and $P(Q=Q_2)=0.5$, then the variance of the distribution is $\mathrm{Var}(Q) = (Q_2 - Q_1)^2 / 4$ and the susceptibility $\chi \equiv N \mathrm{Var}(Q) = N (Q_2 - Q_1)^2 / 4$. This, exactly, is the blue dashed line in Fig.~\ref{fig:X_N} which approximates $\chi^\star_\mathrm{dis}$ decently well. Deviations from that line increase for larger $N$ because the high-overlap peak $Q_2$ shifts to smaller values, see Fig.~\ref{fig:P_Q_Ec_averaged_differentN}e.

%

The apparent proportionality $\chi_\mathrm{dis}^\star \propto N$ has been used to support the idea that RFIM criticality is present in this model~\cite{RFIM_in_glassforming_liquid,guiselin2022statistical}. However, this proportionality seems to be the consequence of bimodality of the overlap distribution, cf.~Fig.~\ref{fig:P_Q_Ec_averaged_differentN}: At the low temperature $T=0.07$, the distribution is bimodal with two widely separated peaks \textit{for the considered system sizes} $N$. Therefore, the disorder fluctuations in the overlap are dominated by those two peaks (according to a Bernoulli distribution discussed above) such that $\chi_\mathrm{dis}^\star \propto N$. However, bimodality seems to be a finite-size effect, as discussed before, so that the proportionality is not granted for sufficiently large systems. Quite the contrary, we expect that both susceptibilities $\chi_\mathrm{dis}^\star$ and $\chi_\mathrm{con}^\star$ are finite for $N \to \infty$: The connected susceptibility $\chi_\mathrm{con}^\star$ measures spatial correlations between particles, which should be of finite extent at any temperature (as numerous studies of dynamical and static length scales demonstrate). 

To better understand the susceptibilities and finite-size effects above, we explicitly determine the spatial distribution of the particle overlaps with a correlation function in the next section. We will see that clusters of low and high overlap grow with decreasing temperature.



\subsection{Growing static length scale \label{sec_ta_d}}
As we saw in Sec.~\ref{sec:relaxation_dynamics}, the coupling of the fluid to a reference configuration via an external field $\varepsilon$ increases its kinetic stability. For critical field strength $\varepsilon^\star$, once the stationary state is reached, \textit{thermodynamic} fluctuations of the particles are observed that grow with decreasing temperature. With their analysis in the previous section, we did not observe a critical point (above the numerical glass-transition temperature, even though the fast swap Monte Carlo algorithm is used). However, the phenomenology of increasing fluctuations unambiguously reveals that particles move increasingly in a collective manner. This is equivalent to the growth of a static length scale. In this section we reveal the growing length scale more directly with a spatial correlation function $C_q$ of local overlaps and illustrate it with snapshots of the liquid.

\textbf{Biasing potential.}
We want to understand how particles $i$ with low and those with high local overlap $q({\bf r}_i)$, see Eq.~(\ref{eq:Q_definition}), spatially arrange. This is only meaningful when the global overlap $Q$ takes on an intermediate value between $Q_\mathrm{rnd} \approx 0$ and $1$. For this purpose we force the system to $Q\approx 0.36$ by applying a harmonic bias potential (as used in umbrella sampling), i.e., the term $k N (Q-b)^2/2$ is added to the Hamiltonian (\ref{eq:H_epsilon_definition}) with $k = 20$ and $b = 0.36$.
After relaxation in the biasing potential, about $50\%$ of the particles have a local overlap $q({\bf r}_i) \approx 1$ and for the other half it is $q({\bf r}_i) \approx 0$. Such configurations are analyzed in this section.

\textbf{Correlation function $C_q(r)$ of the local overlaps.}
With decreasing $T$, particles $i$ organize in clusters of low and others of high local-overlap $q({\bf r}_i)$. We can measure this with the spatial correlation function
\begin{align}
    C_q(r) &= 
    \left\langle \frac{
    \sum_{i<j} {\color{red} q({\bf r}_i) q({\bf r}_j)}\mathbf{1}_{[r -\delta r,\, r+\delta r]}(|{\bf r}_i - {\bf r}_j|)
    }
    {
    \sum_{i<j} \mathbf{1}_{[r -\delta r,\, r+\delta r]}(|{\bf r}_i - {\bf r}_j|)
    } \right\rangle. 
    \label{eq:C_q_r}
\end{align}
%
Here the sums run over all particle pairs $(i,j)$. The indicator function $\mathbf{1}_{[r -\delta r,\, r+\delta r]}(|{\bf r}_i - {\bf r}_j|)$ is $1$ when the distance between particles $i$ and $j$ equals $r$ (except for a tolerance $\delta r = 0.005$), and otherwise $0$.
Thus $C_q(r)$ measures the overlap correlation $\color{red}q({\bf r}_i) q({\bf r}_j)$ between all pairs of particles which are separated by a distance $r$.

\begin{figure}
        \centering
        \includegraphics{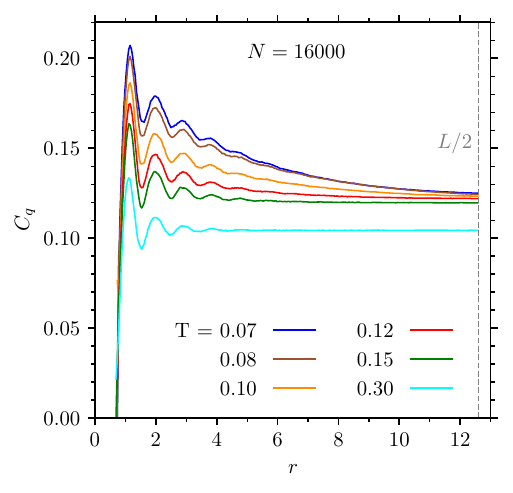}
        \vspace{-0.7cm}
        \caption{Local-overlap correlation function $C_q$ versus particle distance $r$
        when particles are forced to intermediate overlap $Q \approx 0.36$ with a biasing potential. 
        Different temperatures $T$ are considered. 
        The vertical line indicates half the simulation-box length, with $L = N^{1/3}$ and $N=16000$ particles.}
        \label{fig:C_q_r_variousT}
\end{figure}

Figure \ref{fig:C_q_r_variousT} shows 
$C_q$ as a function of $r$ for different temperatures $T$ for a large system, $N=16000$ particles.
All curves show a modulation resembling a local structure as measured by the radial distribution function. 
For $r \to \infty$ there should be no correlation between particles,
\begin{align}
    C_q^\infty 
    &:= \lim_{r \to \infty} C_q(r)
    = \langle Q \rangle^2 \approx b^2 = 0.1296. 
\end{align}
%
%
The unbiased system does not want to be at $Q > Q_\mathrm{rnd}$, so that $C_q^\infty < b^2$. By choosing $\varepsilon \approx \varepsilon^\star(T)$ (or by increasing the spring constant $k$) one can decrease those deviations.

At the highest temperature $T=0.30$ (cyan), $C_q(r)$ is approximately constant for $r \gtrsim 1$, except for the trivial modulation mentioned above.
With decreasing temperature, correlation increases at a fixed $r$. Noticeably a ``long-ranged'' decay develops, i.e., particles tend to have similar overlap values even when displaced by large distances $r$. In this sense, $C_q(r)$ reveals that particles organize in large clusters of high and low local overlap at low temperature.
Our results unambiguously reveal the growth of a static length scale, which might correspond to the point-to-set length measured in glassforming liquids.

\begin{figure}
        \centering
        \hspace{-0.27cm}
        \includegraphics{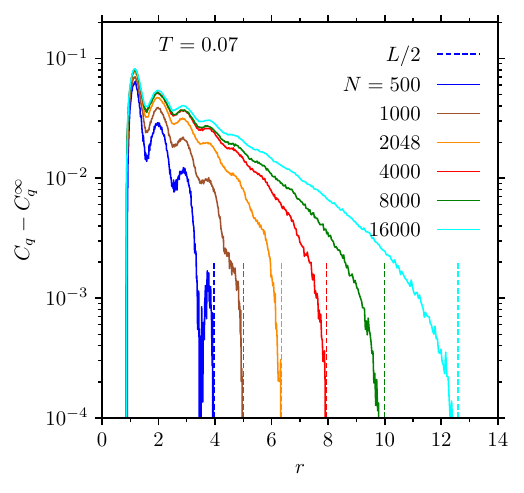}
        \vspace{-0.7cm}
        \caption{Excess part of the local-overlap correlation function, $C_q(r) - C_q(r \to \infty)$, versus particle distance $r$ when the system is biased to $Q\approx 0.36$. 
        Many different system sizes with box lengths $L = N^{1/3}$ are considered at fixed $T =0.07$. In each case, $L/2$ is indicated by a vertical line.}
         \vspace{-0.05cm}
        \label{fig:C_q_r_variousN}
\end{figure}

Figure~\ref{fig:C_q_r_variousN} shows $C_q(r)- C_q^\infty$ as a function of $r$ at $T=0.07$ for many system sizes, $N = 500$--16000 particles. Huge finite-size effects are found, as the curves for small systems decay much faster than for larger ones. $C_q(r)$ shows relatively sharp drops close to half the simulation box, $r \lesssim L/2$. This shows that the size of the clusters is limited by the simulation-box size at $T=0.07$.

\begin{figure*}
\includegraphics[scale=1.1]{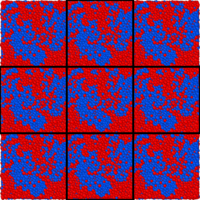}
\includegraphics[scale=1.1]{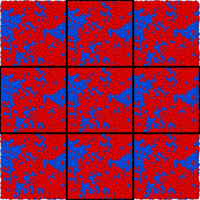}
\includegraphics[scale=1.1]{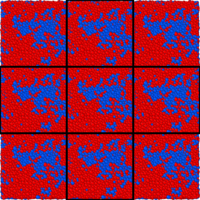}\\
\includegraphics[scale=1.1]{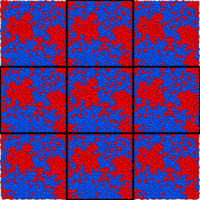}
\includegraphics[scale=1.1]{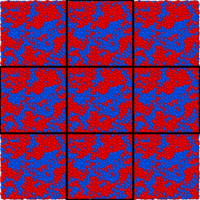}
\includegraphics[scale=1.1]{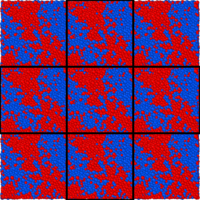}\\
\includegraphics[scale=1.1]{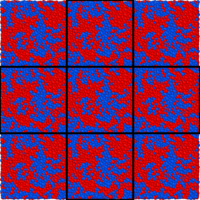}
\includegraphics[scale=1.1]{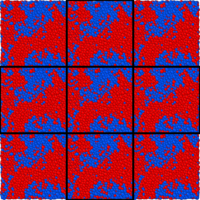}
\includegraphics[scale=1.1]{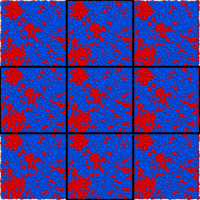}\\
\includegraphics[scale=1.1]{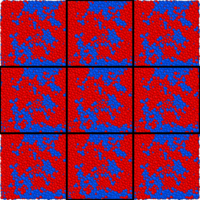}
\includegraphics[scale=1.1]{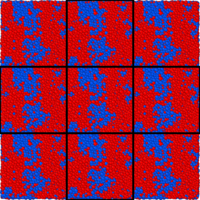}
\includegraphics[scale=1.1]{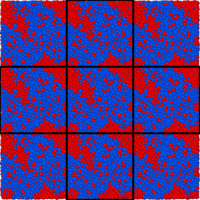}
	\caption{Snapshots of systems with $N=16000$ particles at $T=0.07$ and intermediate overlap $Q \approx 0.36$.
    Particles $i$ with local overlap $q({\bf r}_i) > 0.2132$ are shown in red, the others in blue.
    The original system \textit{and} $8$ periodic images (indicated by black boxes) can be seen. Each row shows the same configuration, but three different views. Plots were created with OVITO \cite{stukowski2009visualization}.}
    \label{Fig:snapshots_intermediate_Q_N16k}
\end{figure*}

\begin{figure*}
\includegraphics[scale=1.10]{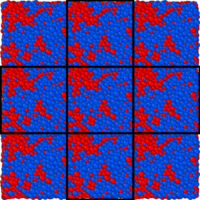}
\includegraphics[scale=1.10]{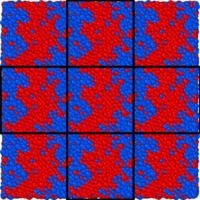}
\includegraphics[scale=1.10]{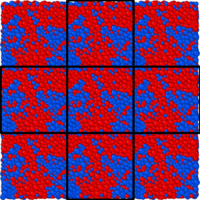}\\
\includegraphics[scale=1.10]{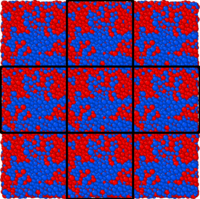}
\includegraphics[scale=1.10]{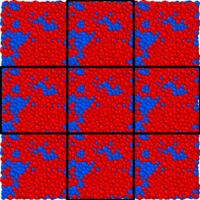}
\includegraphics[scale=1.10]{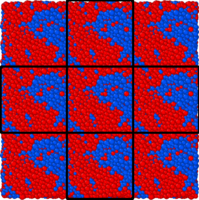}\\
\includegraphics[scale=1.10]{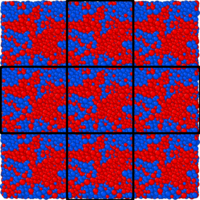}
\includegraphics[scale=1.10]{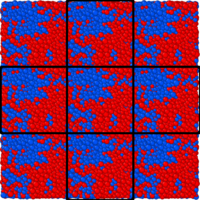}
\includegraphics[scale=1.10]{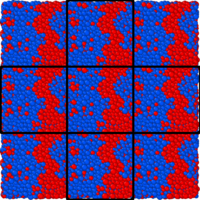}\\
\includegraphics[scale=1.10]{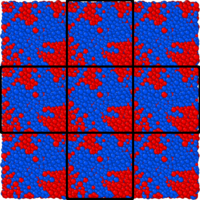}
\includegraphics[scale=1.10]{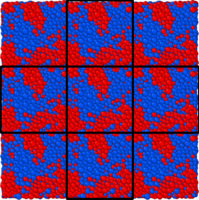}
\includegraphics[scale=1.10]{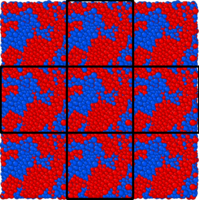}
\caption{Snapshots of systems with $N=4000$ particles at $T=0.07$ and intermediate overlap $Q \approx 0.36$.
    Particles $i$ with local overlap $q({\bf r}_i) > 0.2132$ are shown in red, the others in blue.
    The original system \textit{and} $8$ periodic images (indicated by black boxes) can be seen. Each row shows the same configuration, but three different views. Plots were created with OVITO \cite{stukowski2009visualization}.}
    \label{Fig:snapshots_intermediate_Q_N4k}
\end{figure*}

\begin{figure*}
\includegraphics[scale=1.1]{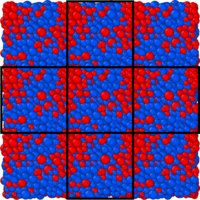}
\includegraphics[scale=1.1]{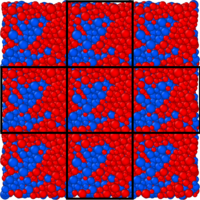}
\includegraphics[scale=1.1]{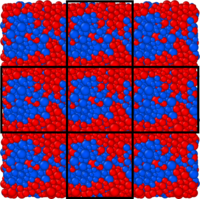}\\
\includegraphics[scale=1.1]{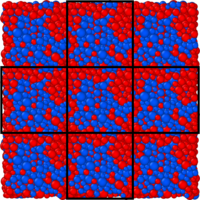}
\includegraphics[scale=1.1]{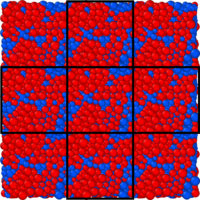}
\includegraphics[scale=1.1]{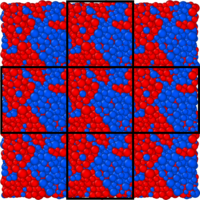}\\
\includegraphics[scale=1.1]{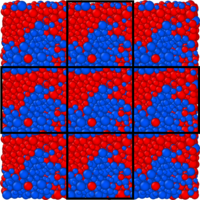}
\includegraphics[scale=1.1]{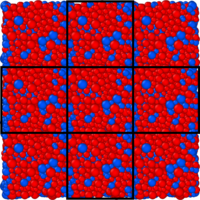}
\includegraphics[scale=1.1]{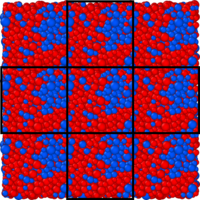}\\
\includegraphics[scale=1.1]{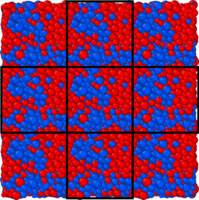}
\includegraphics[scale=1.1]{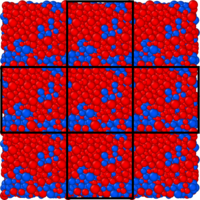}
\includegraphics[scale=1.1]{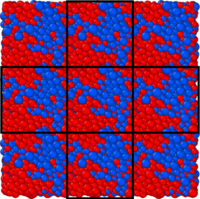}
	\caption{Snapshots of systems with $N=1000$ particles at $T=0.07$ and intermediate overlap $Q \approx 0.36$.
    Particles $i$ with local overlap $q({\bf r}_i) > 0.2132$ are shown in red, the others in blue.
    The original system \textit{and} $8$ periodic images (indicated by black boxes) can be seen. Each row shows the same configuration, but three different views. Plots were created with OVITO \cite{stukowski2009visualization}.}
    \label{Fig:snapshots_intermediate_Q_N1k}
\end{figure*}

To illustrate how the particles organize in clusters at low temperature $T=0.07$ and at intermediate overlap, we show many snapshots of the liquid in Fig.~\ref{Fig:snapshots_intermediate_Q_N16k} ($N=16000$), Fig.~\ref{Fig:snapshots_intermediate_Q_N4k} ($N=4000$), and Fig.~\ref{Fig:snapshots_intermediate_Q_N1k} ($N=1000$). These correspond to configurations analyzed for Fig.~\ref{fig:C_q_r_variousN} as well. Each row represents a different configuration, showing the same configuration from three different views. Each view shows not only the original system but also $8$ (of all $26$) periodic images, indicated by black boxes.
Particles whose local overlap is larger than the median are shown in red, the other half 
in blue (note that the median $\approx 0.2132$ is smaller than the arithmetic mean, the global overlap $Q \approx 0.36$).
We see that particles with high and those with low overlap organize in heterogeneous clusters, whose geometry strongly varies.
For all system sizes the clusters are of a similar size as the simulation box,
consistent with the finite-size effects observed in Fig.~\ref{fig:C_q_r_variousN}.
In contrast, for the liquid at high temperatures $T \geq 0.3$, the particles with overlap are essentially randomly distributed among all particles (not shown here), as it is the case for the ideal gas, see App.~\ref{app:ideal_gas}.



\section{Interpretation and conclusion \label{sec:conclusions}}
We have seen that replica-coupled liquids at low temperature $T$ can organize in clusters of high and low overlap. This is associated with an increase of susceptibilities (fluctuations) and the transition from unimodal to bimodal overlap distributions along a line of field strengths $\varepsilon^\star(T)$ where the fluctuations are maximal. 

Unlike recent simulation studies \cite{berthier2015evidence,RFIM_in_glassforming_liquid,guiselin2022statistical}, we argue below that the bimodal distributions do not indicate first-order phase transitions that occur below a critical temperature. We do not find evidence for such a critical temperature belonging to the universality class of the RFIM, as claimed in those studies. Instead, we argue below that the transition from unimodality to bimodality is a finite-size effect that is the result of a growing static length scale $\xi$. This correlation length measures the spatial extent of correlation between particles, domains of high overlap, as identified in Sec.~\ref{sec_ta_d} with snapshots and a local-overlap correlation function $C_q(r)$.

\textbf{A finite length scale $\xi$.}
For our interpretation we assume that the length scale $\xi(T)$ is always \textit{finite} for any finite temperature $T$. This assumption seems plausible considering the smooth growth of $\xi$ with decreasing $T$ in Sec.~\ref{sec_ta_d} and when drawing the analogy between $\xi$ and the point-to-set length (discussed below). Thus, in the thermodynamic limit $N\to\infty$, we expect that the glassforming liquid  at $\varepsilon^\star$ looks \textit{homogeneous} on length scales much larger than $\xi$ while on length scales of the order or smaller than $\xi$ it appears to be preferably in a state with either high or low overlap. Then, we can explain the development of bimodal distributions as follows.

At high temperatures the particles are merely correlated, such that the correlation length $\xi$ is much smaller than the linear size $L$ of the simulation box. Here the distribution of the overlap is unimodal, because the system is homogeneously constituted of many small clusters of high and low overlap, each of size $\xi \ll L$. This situation is qualitatively similar to the \textit{ideal gas} for which the particles are completely \textit{independent} of each other such that the total overlap $Q$ follows a binomial (single-peak) distribution, as analytically derived in App.~\ref{app:ideal_gas}.

When the temperature is decreased, the correlation length $\xi$ increases. When $\xi$ is of the order of the linear system size, $\xi \sim L$, then the overlap distributions are very broad and start to become bimodal. 
With further decreasing the temperature, $\xi$ becomes much larger than $L$ and the system is preferably found either in a high- or low-overlap state. This can be explained via a sub-box analysis, considering many sub-systems of size $L$ of a macroscopic system: Since the clusters in the homogeneous macroscopic system are of size $\xi \gg L$, most subsystems will either have full overlap or no (i.e., random) overlap. Thus most individual distributions of the subsystems are unimodal, with single peaks at low or high overlap. However, since $\varepsilon = \varepsilon^\star$, there are equally many low and high overlap peaks, such that the disorder-averaged distribution is bimodal as a sum of those peaks.

According to this interpretation, bimodality clearly is a finite-size effect, because one would always observe a unimodal distribution if one were able to consider sufficiently large systems, i.e., $L\gg \xi$, for all temperatures. Our data for overlap distributions at different system sizes indicate that this is indeed the case, cf.~Fig.~\ref{fig:P_Q_Ec_averaged_manyT}.

\textbf{Interfacial free energy.}
At $T=0.07$, where a pronounced bimodal distribution is observed, we used a harmonic bias to force the system to an intermediate value of the overlap ($Q \sim 0.5$), see Sec.~\ref{sec_ta_d}. Here, we see domains of high and low overlap that are separated from each other by interfaces.  Such a state corresponds to a very low probability and thus a high free energy. In this sense, the formation of those interfaces costs free energy for a system of the given size. For a large system with $L\gg \xi$, however, domains of the preferable size $\xi$ can form and thus the overlap distribution is expected to become unimodal also at low temperature. 


\textbf{A mathematical argument.}
The interpretation of the transition from unimodal to bimodal distributions in terms of a finite-size effect can also be understood via a simple mathematical argument on variances:
First note that fluctuations of a bounded random variable $Q \in [0,\,1]$ are bounded by \textit{Popoviciu's inequality on variances}~\cite{popoviciu1935equations} according to $\mathrm{Var}(Q) \leq 1/4$. Equality holds exactly when $Q$ follows a Bernoulli distribution with $P(Q=1) = P(Q=0) = 0.5$, which, of course, has a bimodal shape.
Now consider a system of fixed size $N$ whose fluctuations $\chi := N \mathrm{Var}(Q)$ steadily increase upon variation of a control parameter $T$. When $\chi$ approaches the limiting value $N/4$ given by Popoviciu, a transition to a bimodal distribution \textit{must occur} as it is the only way to increase the fluctuations toward $N/4$.

\textbf{Comparison to point-to-set length.}
There are similarities of replica coupling with particle-pinning methods \cite{montanari2006rigorous,cavagna2007mosaic}, with the difference that in replica coupling the system can choose its own flexible ``frozen'' boundary. We believe that the size of the detected overlap domains, as quantified by $\xi$, corresponds to the point-to-set length scale. At the low temperature $T=0.07$ we can infer a length scale of $\xi \sim 10$ from Fig.~\ref{fig:C_q_r_variousN} (e.g., by defining $C_q(\xi) - C_q^\infty = 0.002$). This relatively large length scale can be compared to point-to-set lengths obtained from other simulation studies where $\xi \sim 2.5$--7 \cite{cavagna2007mosaic,biroli2008thermodynamic,biroli2013comparison,karmakar2015length}. Of course, such a comparison should be taken with a grain of salt due to different definitions. However, that we obtain a larger length scale is probably due to the fact that we reach temperatures that are farther below $T_{\rm MCT}$ than those reached in the previous studies (due to the efficiency of polydispersity with respect to the swap algorithm). In forthcoming studies, point-to-set lengths of our model shall be explicitly determined to clarify the link between replica coupling and particle pinning.

\textbf{Relation to bulk dynamics.}
An open question is whether the growing length scale extracted from the replica-coupled liquid is related to the dynamics of unconstrained liquids below $T_{\rm MCT}$. As discussed in the introduction, for temperatures $T< T_{\rm MCT}$ the liquid is in amorphous solid state on a finite timescale. Here collective rearrangements of particles are required to allow for the decorrelation from the initial configuration at time $t=0$. The spatial extent of these rearrangements, a dynamic length scale $\xi_\mathrm{d}$, can be measured from dynamic susceptibilities \cite{berthier2011theoretical} which describe fluctuations of the overlap $Q$ (with respect to the initial configuration at $t=0$) around the average dynamics. In the bulk system, dynamic susceptibilities are maximal at a finite time around the alpha-relaxation time (where $Q\sim0.5$), at which point $\xi_\mathrm{d}$ is determined. Similarly, with replica coupling the overlap fluctuations are maximal at a critical field strength $\varepsilon^\star$ (where $Q\sim 0.5$ as well), cf.~Secs.~\ref{sec:relaxation_dynamics} and \ref{sec:thermodynamic_analysis}. In this case, however, the susceptibilities are calculated once the stationary state is reached and therefore referred to as static or thermodynamic fluctuations. We think that it is a crucial question for the understanding of the glass transition to elucidate the relation between the correlation length $\xi$ of replica-coupled liquids with the dynamic one, $\xi_\mathrm{d}$, obtained for unconstrained liquids.

\appendix

\section{Random overlap $Q_\mathrm{rnd}$}\label{app:random_overlap}
To compute the random overlap $Q_\mathrm{rnd} := \langle Q \rangle$, we first note that the probability density $P({\bf r}_i)$ to find a particle $i$ at ${\bf r}_i$ is given by $P({\bf r}_i) = 1/V$ where $V$ is the volume of the simulation box $\mathcal{V}$. The global probability density $P_g(\{{\bf r}_k\})$ to find $N$ particles at positions $\{{\bf r}_k\}$ is related to $P$ by integrating out all other degrees of freedom,
%
\begin{equation}
P({\bf r}_i) = \int_{\mathcal{V}^{N-1}} P_g(\{{\bf r}_k\})
\text{d}{\bf r}_1 .. \text{d}{\bf r}_{i-1}  \text{d}{\bf r}_{i+1} .. \text{d}{\bf r}_{N}.
\end{equation}
From Eqs.~(\ref{eq:Q_definition}) and (\ref{eq:overlap_function_continuous}), we calculate
\begin{align}
Q_\mathrm{rnd}  &\equiv \int_{\mathcal{V}^N} Q(\{{\bf r}_i\},\{{\bf r}_j^0\}) P_g(\{{\bf r}_i\}) \text{d}{\bf r}_i^N \\
&= \frac{1}{V} \int_{\mathcal{V}}  \sum_{j=1}^{N} \omega( |{\bf r} - {\bf r}_j^0| /a ) \text{d}{\bf r}\\
&= \frac{N 4 \pi a^3}{V}  \int_{0}^{1} \omega( x ) x^2 \text{d}{x}\\
&= \frac{5}{28} \frac{4 \pi a^3}{3} \rho \approx 0.0479.
\end{align}
In the case of the binary overlap where $\omega(x) = \Theta(1-x)$, we obtain $Q_\mathrm{rnd} = \frac{4 \pi a^3}{3} \rho$, the volume fraction of $N$ spheres with radius $a$ (at positions ${\bf r}_j^0$).


\section{Exact analytical solution of the ideal-gas overlap-distribution in the external field\label{app:ideal_gas}}
For the ideal gas, an \textit{exact} analytical expression for the overlap-distribution $P_\varepsilon(Q)$ in the presence of the $\varepsilon$-field can be derived. This is demonstrated in this section, starting with general definitions that also hold for the liquid. Then the ideal gas is considered by setting the potential energy to zero, $U = 0$. The ideal gas case provides a reference for the understanding of replica-coupled liquids, clarifying for example the Widom line $\varepsilon^\star(T)$. 


\textbf{Partition function.} 
The starting point for our calculations is the partition function $\mathcal{Z}_\varepsilon$ of a macrostate $Q$ in the canonical ensemble, counting all states at a given $Q$ via
\begin{align}
    \mathcal{Z}_\varepsilon(Q|{\bf r}^0) &= \frac{1}{N!} \int\delta\left(Q - \hat{Q}({\bf r},{\bf r}^0)\right) e^{-\beta H_\varepsilon({\bf r}|{\bf r}^0)}\,\text{d}{\bf r} .
    \label{eq:partition_function_Q_1} 
\end{align}
Here, $\hat{Q}$ is the overlap function defined in Eq.~(\ref{eq:Q_definition}), $\delta$ is the Dirac delta distribution, $\beta = (k_\mathrm{B} T)^{-1}$, and $H_\varepsilon$ the Hamiltonian including the external field, Eq.~(\ref{eq:H_epsilon_definition}), which couples the liquid to the reference configuration ${\bf r}^0$. The integral runs over all $3\times N$ coordinates ${\bf r}$ of the liquid, i.e., for each particle over the whole volume of the simulation box. In the case that the \textit{local} overlaps $q$ are binary the \textit{total} overlap $Q = k/N$ is a rational number, such that $\delta$ becomes the Kronecker delta $\delta_{Q\hat{Q}}$ instead. The partition function $\mathcal{Z}_\varepsilon$ measures the number of available microstates at a specific value $Q$ of the overlap, each weighted by a Boltzmann factor $e^{-\beta H_\varepsilon}$. The division by $N!$ ensures that the free energy density does not depend on $N$ for $N \to \infty$.

Analogue to $\mathcal{Z}_\varepsilon$, we define the \textit{total} partition function by
\begin{align}
    S_\varepsilon({\bf r}^0) = \frac{1}{N!} \int e^{-\beta H_\varepsilon({\bf r}|{\bf r}^0)}\,\text{d}{\bf r}
    \equiv \int_0^1  \mathcal{Z}_\varepsilon(Q|{\bf r}^0) \,\text{d}Q.
    \label{eq:total_partition_function_2}
\end{align}

\textbf{Probability density.}
The overlap distribution can be defined from the partition functions as
\begin{align}
    P_\varepsilon(Q|{\bf r}^0) = \frac{\mathcal{Z}_\varepsilon(Q|{\bf r}^0)}{S_\varepsilon({\bf r}^0)}.
    \label{eq:P_Q_derivation}
\end{align}
%


\textbf{Partition function of the ideal gas.}
For the ideal gas with binary local overlaps, i.e.~$q({\bf r}_i) \in \{ 0, 1\}$, the partition function $\mathcal{Z}_\varepsilon$ can be calculated analytically. In this case, the window function is given by $\omega(x) = \Theta(1-x)$ and we have to assume that in the reference configuration ${\bf r}^0$ all particles have a mutual distance larger than the microscopic distance $a$ (as otherwise there were positions with $q({\bf r}_i) > 1$). While this assumption holds for typical liquid configurations for temperatures $T \leq 0.3$, it can be introduced as an approximation for the low-density ideal gas. Then, the partition function for the ideal gas is
%
\begin{align}
    \mathcal{Z}_\varepsilon^{\mathrm{ig}} &=  \frac{1}{N!} e^{\beta \varepsilon Q N} \int \delta_{Q,\hat{Q}({\bf r}|{\bf r}^0)} \,d{\bf r} .
    \label{eq:partition_function_Q_idealgas}
\end{align}
Here, $\delta$ is the Kronecker delta, as $Q = k/N$ is always a rational number. The natural number $k \in \{0,\,1,\,\dots,\,N\}$ denotes the number of particles $i$ with overlap $q({\bf r}_i) = 1$.
We can explicitly calculate the integral in Eq.~(\ref{eq:partition_function_Q_idealgas}). Note the similarity to a binomial distribution:
The integrand is only non-trivial when exactly $k = N Q$ particles have local overlap with a reference particle. Thus the integral splits into a sum of $\binom{N}{k} = \frac{N!}{k! (N-k)!}$ non-trivial terms according to the possibilities to have $k$ of $N$ particles with overlap.
Within the simulation box, the regions where a tagged particle has overlap is given by $N$ disjoint spheres, each of volume $V_a = (4/3)\pi a^3$. We obtain
\begin{equation}
    \mathcal{Z}_\varepsilon^{\mathrm{ig}} = \frac{1}{N!} e^{\beta \varepsilon k} \binom{N}{k} (N V_a) ^{k} (V - N V_a)^{N-k}.
    \label{eq:partition_function_Q_idealgas2}
\end{equation}
This expression can be simplified by introducing the volume fraction $p := N V_a / V$ of regions with overlap and the volume $V$ of the whole simulation box. Note that for \textit{binary} local overlap, as assumed for this section, $p \equiv Q_\mathrm{rnd}$ corresponds to the random overlap, cf.~App.~\ref{app:random_overlap}. Then
\begin{equation}
    \mathcal{Z}_\varepsilon^{\mathrm{ig}} = \frac{V^N}{N!} e^{\beta \varepsilon k} \binom{N}{k} p^{k} (1-p)^{N-k}.
    \label{eq:partition_function_Q_idealgas3}
\end{equation}
In the case $\varepsilon=0$, we can identify a binomial distribution $P_0^{\mathrm{ig}}(Q=k/N) = \binom{N}{k} p^{k} (1-p)^{N-k}$.
This is a plausible result, because \textit{for an ideal gas} the term $Q N = \sum_i^N q({\bf r}_i)$ is a sum of $N$ \textit{independent} Bernoulli variables.

For the general case, $\varepsilon \geq 0$, the trick is to rewrite Eq.~(\ref{eq:partition_function_Q_idealgas3}) such that it corresponds to another binomial distribution, however with a different parameter. It is
\begin{align}
    \mathcal{Z}_\varepsilon^{\mathrm{ig}} 
    = \frac{V^N}{N!} \binom{N}{k} \left(\frac{p e^{\beta \varepsilon}}{1-p}\right)^k (1-p)^N . \label{eq:partition_function_Q_idealgas4}
\end{align}
For the term raised to the power of $k$ we have
\begin{align}
  \frac{p e^{\beta \varepsilon}}{1-p}
  = 
  \underbrace{\frac{ e^{\beta \varepsilon}}{ e^{\beta \varepsilon}+ \frac{1-p}{p}}}_{=: p_\varepsilon} \underbrace{ \frac{ e^{\beta \varepsilon} + \frac{1-p}{p}}{\frac{1-p}{p}} }_{\equiv (1 - p_\varepsilon)^{-1}}.
\end{align}
With this definition of $p_\varepsilon$, we get our final result
\begin{align}
    \mathcal{Z}_\varepsilon^{\mathrm{ig}} 
    &=  \frac{V^N}{N!} \binom{N}{k}  p_\varepsilon^k (1-p_\varepsilon)^{N-k} \left( \frac{1-p}{1-p_\varepsilon}\right)^N.
\end{align}
Summing over $k$ yields the total partition function
\begin{align}
    S_\varepsilon^{\mathrm{ig}} = \frac{V^N}{N!} \left( \frac{1-p}{1-p_\varepsilon}\right)^N
\end{align}
such that the overlap distribution is
\begin{equation}
    P_\varepsilon^{\mathrm{ig}}(Q) = \binom{N}{k} p_\varepsilon^{k} (1-p_\varepsilon)^{N-k}
    \label{eq:P_Q_eps_idealgas}
\end{equation}
with $Q = k/N$ and with parameter
\begin{equation}
    p_\varepsilon = p \frac{e^{\beta \varepsilon}}{1 + p(e^{\beta \varepsilon}-1)}.
    \label{eq:critical_epsilon_idealgas}
\end{equation}
%

%
\textbf{Discussion \& interpretation.}
The binomial distribution $P_\varepsilon^{\mathrm{ig}}$ is a unimodal distribution with a single peak at $Q = p_\varepsilon$, the expectation value of $Q$. The binomial distribution converges to a Gaussian distribution for $N \to \infty$ according to the \textit{de Moivre–Laplace limit theorem}. By increasing $\varepsilon$, the peak can be arbitrarily tuned between $p_{\varepsilon=0} = p$ and $\lim_{\varepsilon \to \infty} p_\varepsilon = 1$. This reflects how particles are increasingly trapped in potential wells of depth $\varepsilon$, compensating for the small probability $p \equiv Q_\mathrm{rnd} \propto a^3$ to find a particle at a position with overlap in the absence of the field. Furthermore, we see that $P_\varepsilon^{\mathrm{ig}}$ does not depend on the reference configuration ${\bf r}^0$; except for our initial assumption that particles in ${\bf r}^0$ do not overlap to ensure binary overlap. Therefore, quenched disorder is absent in the ideal gas, as expected.

The variance of the binomial distribution (i.e., of $QN$) is $N p_\varepsilon (1 - p_\varepsilon)$ such that the variance of $Q$ is
\begin{equation}
    \mathrm{Var}_\varepsilon^{\mathrm{ig}}(Q) = \frac{1}{N} p_\varepsilon (1 - p_\varepsilon). \label{eq:Var_Q_ig}
\end{equation}
The width of the distribution $P_\varepsilon^{\mathrm{ig}}$ shrinks for $N \to \infty$ such that its peak becomes increasingly sharp. 
%

\textbf{Widom line $\varepsilon^\star(T)$.} 
For the case of the replica-coupled liquid, we defined a critical $\varepsilon^\star$, cf.~Eq.~(\ref{eq:critical_epsilon}), as the field strength for which fluctuations $\chi = N \mathrm{Var}(Q)$ are maximal. We can explicitly calculate the ``Widom line'' $\varepsilon^\star(T)$ for the ideal gas: For a binomial distribution the variance (\ref{eq:Var_Q_ig}) has a well-defined maximum at $p_{\varepsilon} = 0.5$. Thus the critical $\varepsilon^\star$ can be obtained from Eq.~(\ref{eq:critical_epsilon_idealgas}), yielding
\begin{equation}
    \varepsilon^\star(T) = k_\mathrm{B}T \ln\left(\frac{1-p}{p}\right). \label{eq:widom_line_idealgas}
\end{equation}
To obtain maximal fluctuations, the $\varepsilon$-field has to compensate for the small random overlap $p \equiv Q_\mathrm{rnd} \propto a^3 \approx 0$. 
We see that the critical $\varepsilon^\star$ scales with the thermal energy, $\varepsilon^\star \propto k_\mathrm{B} T$, showing that particles are more likely to escape the potential wells of constant depth $\varepsilon$ when the temperature $T$ is increased. 
How does the Widom line (\ref{eq:widom_line_idealgas}) qualitatively compare to that of the liquid in Fig.~(\ref{fig:Ec_T})? In the $\varepsilon/(k_\mathrm{B}T)$-versus-$T$ diagram, the ideal-gas Widom-line corresponds to a horizontal line. With increasing $T$ the fluid can be expected to behave increasingly similar to the ideal gas. Truly, the beginning of a saturation of the fluid Widom-line can be observed. A quantitative comparison is difficult because we assumed binary overlap in this section, but we used a smooth window function in our simulations of the liquid. Nonetheless, inserting the value $p = Q_\mathrm{rnd} \approx 0.0478$ into Eq.~(\ref{eq:widom_line_idealgas}) yields a surprisingly good result with $\varepsilon^\star/(k_\mathrm{B}T) \approx 2.99$.


\textbf{Free energy.}
With the partition function $\mathcal{Z}_\varepsilon(Q)$, we can define the Boltzmann free energy at overlap $Q$ as
\begin{align}
F_\varepsilon(Q|{\bf r}^0)  &= - k_\mathrm{B} T \ln \left(  \mathcal{Z}_\varepsilon(Q) \right). \label{eq:F_Q_definition}
\end{align}
We explicitly calculate $F_\varepsilon$ when using the umbrella-sampling technique, cf.~App.~\ref{app:umbrella_sampling_technique}, in order to obtain the distribution via $ P = C e^{-\beta F_\varepsilon} $ with normalization $C$. By averaging Eq.~(\ref{eq:F_Q_definition}) over the disorder, one obtains the Franz-Parisi potential $\overline{F_\varepsilon(Q|{\bf r}^0)}$.

The free energy $F_\varepsilon$ can be written as
\begin{equation}
    F_\varepsilon(Q) = -\varepsilon N Q + F_0^\mathrm{id}(Q) + F_0^\mathrm{ex}(Q)
\end{equation}
where $F_0^\mathrm{id}$ is the free energy of the ideal gas and $F_0^\mathrm{ex}(Q) = - k_\mathrm{B} T \ln \left( \mathcal{Z}_0/\mathcal{Z}_0^{\mathrm{ig}} \right)$ is the excess free-energy of the liquid, both in the absence of the field.

\textbf{Free energy of ideal gas.}
Using the Stirling formula, $\ln(N!) = N \ln(N) - N + \mathcal{O}(\ln(N))$, the ideal-gas free-energy $F_\varepsilon^\mathrm{id}$ can be calculated. A straightforward calculation yields
\begin{align}
   \frac{F_\varepsilon^\mathrm{id}(Q)}{N k_\mathrm{B} T} =&~(1-Q) \ln(1-Q) + Q \ln \left( \frac{1-p}{p} \right) \nonumber\\
    &+  Q\ln Q - \ln(1-p) +\textit{Const} \label{eq:F_Q_id},
\end{align} 
except for a term of the order of $\mathcal{O}(\ln(N)/N)$ which goes to zero for $N\to\infty$.
The constant is $\textit{Const} = \ln( \rho ) -1$ with $\rho = N / V = 1$. Here, the normalization of the partition function by $N!$ is important; otherwise, it was $\textit{Const} = \ln( \rho ) - \ln(N)$ and the free-energy density were $N$-dependent in the thermodynamic limit.

\section{Umbrella-sampling technique}\label{app:umbrella_sampling_technique}
Here, we describe the umbrella-sampling (US) technique that we use to obtain the equilibrium distribution $P_\varepsilon(Q | {\bf r}_0)$. The idea of US is to force the liquid toward improbable phase-space points by the introduction of a biasing potential in the Hamilton function (see below). With US, one may overcome sampling problems that one has when measuring $P_\varepsilon$ directly via the calculation of a histogram of $Q$ values (after reaching equilibrium in the presence of the external field $\varepsilon$). In direct sampling, improbable values of $Q$ are sampled rarely and thus the ``tails'' of the histogram suffer from poor statistics. Moreover, at low temperature, \textit{in the presence of the $\varepsilon$-field} the relaxation dynamics becomes particularly slow such that even the use of swap dynamics does not allow to reach the steady state which is necessary for correct equilibrium sampling. US also helps to approach steady states that are not accessible via direct sampling. Below, we introduce the US technique and give the main details of our simulations with US. For a review of US see Ref.~\cite{umbrella_sampling_kastner2011}.

\textbf{The biasing potential.}
To force the replica-coupled liquid to states with an overlap $Q$ that are associated with a low probability, a biasing potential $B$ is added to the unbiased Hamilton function $H_0$. Thus the biased Hamilton function is $H^b = H_0 + B$. The potential $B$ depends on $Q$ and pushes the liquid towards the value $b$ of the overlap with the reference configuration. To achieve this, we use a harmonic shape for the biasing potential, $B = N k (Q-b)^2 / 2$, with spring constant $k = 20$ (note that such harmonic potentials are commonly used in US). To probe the whole phase space via $Q$, many independent simulations are carried out, here a total number of $I=43$, each with a different bias $b_i$ given by the list $\{b_i\,|\,i = 1,\dots,I \} = 10^{-2} \times \{ 0$, $1$, $2$, $3$, $4$, $5$, $6$, $7$, $8$, $10$, $12$, $14$, $18$, $20$, $22$, $24$, $26$, $28$, $30$, $32$, $34$, $36$, $38$, $40$, $42$, $44$, $46$, $48$, $50$, $52$, $56$, $58$, $60$, $62$, $64$, $66$, $68$, $70$, $75$, $80$, $85$, $90$, $100 \}$, with more points around the random overlap $Q_\mathrm{rnd} \approx 4.8 \times 10^{-2}$ than for larger $Q$.


\textbf{Sampling in a window.}
In each ``window'' $i$, only a \textit{biased} distribution $P^b_i(Q)$ can be measured. We determine a time series $Q(t)$ of $5000$ equidistant data points, from which only those \textit{equilibrated} under the biasing potential are used: When the bias is switched on at time $t=0$, the system relaxes toward $Q \approx b$. We only use $Q(t)$ for which $0.5\, t_\mathrm{max} < t <  t_\mathrm{max}$ where $t_\mathrm{max} = 10^5 \gg \tau$ is much larger than the bulk relaxation-time $\tau$ of the liquid simulated with fast swap dynamics. To ensure that equilibrium is actually reached, we check whether two different protocols (cf.~Sec.~\ref{sec:relaxation_dynamics}) yield identical results: For the first protocol, we use two independent configurations ${\bf r}$ and ${\bf r}^0$ such that $Q(t=0) \approx Q_\mathrm{rnd}$ with a small random overlap. For the other protocol, we use ${\bf r} = {\bf r}^0$ such that $Q(t=0) = 1$. After switching on the bias, for both protocols, $Q(t)$ approaches and finally fluctuates around an average value $Q = \bar{Q}^b_i$ (note that $\bar{Q}^b_i \approx b_i$ but $\bar{Q}^b_i \neq b_i$). For our variant of umbrella integration discussed below (which uses spline interpolation), we only need very few information from each simulation window: (i) the value of the bias $b_i$ and (ii) the time average $\bar{Q}^b_i$ of the corresponding $Q(t)$ time series.

\textbf{Unbiased and biased distributions.}
Now the key idea of US is that the \textit{unbiased} distribution $P$ can be calculated from the many \textit{biased} distributions $P^b_i$. The biased distribution $P^b_i$ of the overlap $Q$ can be analytically expressed via the canonical distribution of the phase-space coordinates, cf.~Eq.~(\ref{eq:P_NVT_unbiased}) and Eq.~(\ref{eq:P_eps_Q_r0}) where the Hamiltonian is replaced with $H^b$ here. A straightforward calculation relates $P^b_i$ to the unbiased distribution $P(Q)$ via
\begin{equation}
  P^b_i(Q) = C_i P(Q) \exp{\left(-\frac{B_i(Q)}{k_\mathrm{B}T}\right)},
\label{eq:P_Q_reweighting}
\end{equation}
where $C_i$ is a window-dependent constant.
    
For US, it is convenient to define the free energy
\begin{equation}
   F(Q) = -k_\mathrm{B}T \ln{P(Q)} \label{eq:F_Q_definition}
\end{equation}
and calculate
\begin{align}
F(Q) = -k_\mathrm{B}T \ln{P^b_i(Q)} - B_i(Q) + \tilde{C}_i
\label{eq:F_Q_umbrella}
\end{align}
with a new constant $\tilde{C}_i$. The derivation here is analytically exact, but only quantitatively accurate for each window $i$ around values $Q \approx \bar{Q}^b_i$ where the sampling is sufficient.

\textbf{Combining window data: umbrella integration.}
To calculate the unbiased distribution (or equivalently free energy) from the window data, we use a method called \textit{umbrella integration} \cite{umbrella_sampling_kastner2011}. First we take the derivative of Eq.~(\ref{eq:F_Q_umbrella}) to obtain
\begin{equation}
\frac{\partial F}{\partial Q} = -k_\mathrm{B}T \frac{\partial\ln{P^b_i}}{\partial Q} - \frac{\partial B_i}{\partial Q}, 
\label{eq:F_Q_derivative_umbrella}
\end{equation}
which does not depend on $\tilde{C}_i$. Now, for the first time, we approximate the distribution $P^b_i$ with a normal distribution around the average value $Q = \bar{Q}^b_i$:
\begin{equation}
    P^b_i \approx \frac{1}{\sigma_i^b \sqrt{2 \pi}} \exp{\left(-\frac{1}{2 (\sigma_i^b)^2}(Q-\bar{Q}^b_i)^2 \right)}.
\end{equation}
Here, $\sigma_i^b$ is the standard deviation of the $Q$ data in window $i$. This approximation is well justified because each window should cover only a small part of the $Q$-range. With this approximation and the explicit harmonic form of the bias potential $B_i$, Eq.~(\ref{eq:F_Q_derivative_umbrella}) reads
\begin{align}
\frac{\partial F}{\partial Q} = -k_\mathrm{B}T \frac{Q -\bar{Q}^b_i }{(\sigma_i^b)^2} - N k (Q - b_i).
\end{align}
Evaluating this expression at $Q = \bar{Q}^b_i $ yields
\begin{equation}
   \frac{\partial F}{\partial Q}|_{\bar{Q}^b_i} = - N k (\bar{Q}^b_i - b_i).
\end{equation}
We use a cubic spline interpolation on the small discrete set $\{ (\bar{Q}^b_i, \frac{\partial F}{\partial Q}|_{\bar{Q}^b_i})\,|\,i = 1,\dots,I \}$ to obtain the derivative $\frac{\partial F}{\partial Q}$ with a high resolution. Then, we numerically integrate $\frac{\partial F}{\partial Q}$ via Simpson's rule, yielding $F(Q)$ except for an arbitrary integration constant. From $F$ we get back the unbiased distribution $P$ from the definition (\ref{eq:F_Q_definition}). For any value of $\varepsilon$, we obtain the distribution $P_\varepsilon$ from $P$ by re-weighting Eq.~(\ref{eq:P_eps_Q_r0}), analogue to Eq.~(\ref{eq:P_Q_reweighting}). Note that $P_\varepsilon(Q|{\bf r}_0)$ depends on the reference configuration ${\bf r}_0$.  To calculate disorder averages $\overline{(.)}$ (e.g., for the susceptibilities) we repeat the US protocol $10$--$30$ times with different ${\bf r}_0$.
 
\textbf{Discussion of the umbrella-sampling technique.}
A huge advantage of US over the direct-sampling approach is that $P_\varepsilon$ for any $\varepsilon$ can be readily obtained from US without external field ($\varepsilon=0$) by simple re-weighting. This is not possible with the direct method because its sampling in the tails of the distribution is too poor. The downside of US is a high computational cost, because a separate simulation run for each window is necessary.
    


%
\section{Relation between susceptibilities\label{app:relation_between_chis}}
Here, we elucidate the relation between the disconnected and the connected susceptibility. As Fig.~\ref{fig:Xdis_X_con} indicates, at the critical field strength $\varepsilon^\star$, see Eqs.~(\ref{eq:chi_con})--(\ref{eq:critical_epsilon}), the susceptibilities follow $\chi_\mathrm{dis}^\star \propto (\chi_\mathrm{con}^\star)^2$. In the following, we derive the relation
\begin{equation}
    \chi_\mathrm{dis}^\star = N \overline{(\varepsilon^\star - \varepsilon_0)^2} (k_\mathrm{B} T)^{-2} (\chi_\mathrm{con}^\star)^2 \, ,
    \label{eq:chis_relation}
\end{equation}
where the field strength $\varepsilon_0$ is associated with a given reference configuration ${\bf r}_0$, defined as the point where the thermal average of the overlap is equal to the disorder-averaged overlap at critical $\varepsilon^\star$, i.e., $\langle Q|{\bf r}_0 \rangle(\varepsilon_0) = \overline{\langle Q \rangle}(\varepsilon^\star)$.


A Taylor expansion of $\langle Q|{\bf r}_0 \rangle(\varepsilon)$ around $\varepsilon_0$ yields
\begin{equation}
    \langle Q \rangle(\varepsilon) = \langle Q \rangle(\varepsilon_0) + (\varepsilon-\varepsilon_0) \frac{\partial \langle Q \rangle}{\partial \varepsilon}(\varepsilon_0) + \mathcal{O}((\varepsilon-\varepsilon_0)^2).
    \label{eq:Q_thm_taylor}
\end{equation}
For the following, We only need one assumption: The overlap $Q$ is the arithmetic mean of the local overlaps of all $N$ particles. Therefore it should have ``self-averaging behavior'' such that the individual curves $\langle Q \rangle$ approach the disorder average $\overline{\langle Q \rangle}$ for $N \to \infty$. Thus, according to the definition of $\varepsilon_0$ above, we expect $\frac{\partial \langle Q \rangle}{\partial \varepsilon}(\varepsilon_0) \approx \frac{\partial \overline{\langle Q \rangle}}{\partial \varepsilon}(\varepsilon^\star)$ and that $\varepsilon_0 \to \varepsilon^\star$ for $N \to \infty$. Then, evaluating Eq.~(\ref{eq:Q_thm_taylor}) at $\varepsilon^\star$ yields
\begin{equation}
    \langle Q \rangle(\varepsilon^\star) \approx \overline{\langle Q \rangle}(\varepsilon^\star) + (\varepsilon^\star - \varepsilon_0) \frac{\partial \overline{\langle Q \rangle}}{\partial \varepsilon}(\varepsilon^\star).
    \label{eq:Q_approximation}
\end{equation}
This approximation implies $\overline{\varepsilon_0} = \varepsilon^\star$, as we would expect.

In the canonical ensemble the exact relation 
\begin{equation}
   \frac{\partial \langle Q \rangle}{\partial \varepsilon} 
   = N (k_\mathrm{B} T)^{-1} [ \langle Q^2 \rangle - \langle Q \rangle^2] 
\end{equation}
holds. Disorder-averaging yields, by definition,
\begin{equation}
      \frac{\partial \overline{\langle Q \rangle}}{\partial \varepsilon} = 
      (k_\mathrm{B} T)^{-1} \chi_\mathrm{con}.
      \label{eq:Q_derivative_chi_con}
\end{equation}
With Eqs.~(\ref{eq:Q_approximation}) and (\ref{eq:Q_derivative_chi_con}), we obtain
\begin{align} 
     N [\langle Q \rangle(\varepsilon^\star) - \overline{\langle Q \rangle}(\varepsilon^\star)]^2
    = N \frac{ (\varepsilon^\star - \varepsilon_0)^2 }{(k_\mathrm{B} T)^2} (\chi_\mathrm{con}^\star)^2.
\end{align}
Disorder-averaging gives our final result, Eq.~(\ref{eq:chis_relation}). Since $\overline{\varepsilon_0} = \varepsilon^\star$, the term $\mathrm{Var}(\varepsilon_0) = \overline{(\varepsilon^\star - \varepsilon_0)^2}$ denotes the variance of $\varepsilon_0$. We expect $\mathrm{Var}(\varepsilon_0) \propto N^{-1}$, such that both susceptibilities are finite in the thermodynamic limit in the case that either susceptibility is finite.


\bibliography{biblio}
\end{document}